\documentclass[aps,pra,twocolumn,superscriptaddress,notitlepage,showpacs,showkeys]{revtex4-1}
\usepackage{graphicx,amsmath,amsfonts,amssymb,upgreek,txfonts,color}
\usepackage[colorlinks,linkcolor=blue,citecolor=blue,urlcolor=blue,breaklinks=true]{hyperref}
\usepackage{color, colortbl}
\definecolor{lightgreen}{rgb}{0.88,1,1}
\usepackage[left=	2cm,top=2cm,right=1.2cm,bottom=2cm]{geometry}
\usepackage{subcaption}
\usepackage{braket}   
\usepackage{bm,color,graphicx,amsmath,txfonts} 
\newcommand{\be}{\begin{equation}}
	\newcommand{\ee}{\end{equation}}
\newcommand{\bae}{\begin{eqnarray}} \newcommand{\eae}{\end{eqnarray}}

\def\Tr{{\rm tr}}

\begin{document}
	


\title{Quantum thermodynamics, quantum correlations and quantum coherence in accelerating Unruh-DeWitt detectors in both steady and dynamical state}

	\author{Omar Bachain}
	\address{LPHE-Modeling and Simulation, Faculty of Sciences, Mohammed V University in Rabat, Rabat, Morocco}

	\author{Mohamed \surname{Amazioug} }
	\email{m.amazioug@uiz.ac.ma (Corresponding author)}
	\address{LPTHE-Department of Physics, Faculty of Sciences, Ibnou Zohr University, Agadir 80000, Morocco}
	
	\author{Rachid Ahl Laamara}
	\address{LPHE-Modeling and Simulation, Faculty of Sciences, Mohammed V University in Rabat, Rabat, Morocco}
	\address{Centre of Physics and Mathematics, CPM, Faculty of Sciences, Mohammed V University in Rabat, Rabat, Morocco}
	
	\author{Kottakkaran Sooppy Nisar} 
\affiliation{Department of Mathematics, College of Science and Humanities in Al-Kharj, Prince Sattam bin Abdulaziz University, Al-Kharj 11942, Saudi Arabia}
\affiliation{Hourani Center for Applied Scientific Research, Al-Ahliyya Amman University, Amman, Jordan}

\author{Mohammed Zakarya} 
\affiliation{Department of Mathematics, College of Science, King Khalid University, P.O. Box 9004, Abha 61413, Saudi Arabia}

\author{Gamal M. Ismail} 
\affiliation{Department of Mathematics, Faculty of Science, Islamic University of Madinah, Madinah 42351, Saudi Arabia}

\author{Abdel-Haleem Abdel-Aty} 
\affiliation{Department of Physics, College of Sciences, University of Bisha, Bisha 61922, Saudi Arabia}
	\date{\today}
	
\begin{abstract}
	
We investigate the interplay between quantum thermodynamics, quantum correlations, and quantum coherence within the framework of the Unruh-DeWitt (UdW) detector model. By analyzing both the steady and dynamical states of various quantum resources-including steerability, entanglement, quantum discord, and coherence-we study how these resources evolve under Markovian and non-Markovian environments. Furthermore, the hierarchical structure relating quantum correlations and quantum coherence is established. We also examine the thermodynamic performance of a quantum heat engine, highlighting the influence of memory effects and classical correlations on heat exchange, work extraction, and efficiency.
	
\end{abstract}
	
	\keywords{Quantum correlations, Unruh-DeWitt detector, Quantum thermodynamics, Quantum heat engines, Markovian, Non-Markovian, Entanglement, Quantum coherence.}
	
	\maketitle

	\section{Introduction}
	In 1935, Einstein, Podolsky, and Rosen (EPR) introduced a thought experiment, known as the EPR paradox, which posed fundamental challenges to the foundations of quantum mechanics. This paradox revealed the existence of nonlocal correlations between quantum systems, indicating that a measurement performed on one system can instantaneously affect the state of another, spatially separated system \cite{EPR1935,Schrodinger1935}. Bell’s theorem later demonstrated that local hidden variable theories are incapable of accounting for all nonlocal correlations arising from local quantum measurements on spatially separated systems \cite{Bell1964}.
	
	Within this framework, quantum steering has emerged as a particularly intriguing phenomenon in quantum mechanics \cite{Wiseman2007, Uola2020}, highlighting intricate correlations between entangled particles \cite{Cavalcanti2009}. Quantum steering occurs when the measurement of one particle (commonly referred to as Alice) influences the state of another particle (Bob), irrespective of the spatial separation between them, reflecting the inherent nonlocality of quantum mechanics. Leveraging their shared entanglement, Alice can effectively "steer" Bob’s state \cite{Schneeloch2013}.
	
	Beyond entanglement, recent studies have demonstrated that quantum correlations can persist even in separable states \cite{Ferraro2010, Modi2012, Dakic2010}. In this context, geometric quantum discord (GQD) provides an extended framework for characterizing nonclassical correlations beyond entanglement \cite{Hu2018}, offering a powerful tool for quantifying quantum resources. Geometric approaches, such as the Schatten 1-norm quantum discord, have proven particularly useful in a variety of systems, including metal complexes \cite{Cruz2022}. These developments refine our understanding of quantum resource quantification and further expand the potential applications of quantum information protocols \cite{Zeilinger1998, Zeilinger2017Foundations}.
	
	Quantum coherence, a central concept in quantum physics, refers to the capacity of a system to exist in superposition states in which multiple configurations coexist simultaneously. This property underlies a wide range of quintessentially quantum phenomena, including interference and quantum optical effects \cite{Glauber1963,Sudarshan1963}. However, coherence is highly sensitive to environmental disturbances and is progressively suppressed through decoherence \cite{Zurek2003}, a mechanism by which quantum systems lose their non-classical features due to interactions with their surroundings. The interplay between coherence and decoherence plays a crucial role in several domains, notably quantum thermodynamics \cite{Kammerlander2016,Zhang2022}, where it influences the performance and efficiency of quantum heat engines, and the physics of quantum dots \cite{Filgueiras2020,Berrada2020}, which constitute key building blocks for quantum information technologies.
	
	A quantum heat engine (QHE) represents a fundamental platform for investigating problems in quantum thermodynamics (QT), a rapidly developing field that examines thermodynamic processes within the framework of quantum dynamics \cite{Partovi1989,Gemmer2009,UtrerasDiaz2015,RoBnagel2014}. Building upon classical thermodynamic cycles, researchers have proposed their quantum counterparts, enabling the systematic study of QHEs and their efficiencies \cite{Quan2007,Aydiner2018,Cruz2016,Wang2009_PRE,Watanabe2020,Huang2014,Cakmak2017,Su2016,Carnot1824,Kosloff2014}.
	The physical characteristics of QHEs, as well as quantum refrigerators, have been extensively explored in the literature. Typically, a QHE consists of a working substance coupled to two distinct thermal reservoirs—a hot bath and a cold bath. To rigorously analyze the operation of QHEs, both isothermal and isochoric processes are employed. Various quantum systems have been considered as working substances, including single spins \cite{Feldmann2004}, harmonic oscillators \cite{Rezek2006,Lin2003,Insinga2016,Kosloff2017,Oladimeji2019,Oladimeji2021}, ideal quantum gases \cite{Scully2002,Wang2015}, Heisenberg spin models \cite{Huang2013,Peng2019}, and two- or three-level systems \cite{Kieu2004,Quan2005,Wang2012_PRE}. More recently, alternative working materials such as quantum dots \cite{Josefsson2018,Liu2013} and Heisenberg XX, XXZ, or XYZ spin models with Dzyaloshinskii-Moriya interactions (DMI) \cite{He2012,BahaminPili2023_EPJPlus,BahaminPili2023_IJTP,elghaayda2024distribution} have also been investigated, highlighting the growing diversity of quantum systems utilized in heat engine designs. 
	
	Quantum correlations, such as entanglement and quantum discord, play a central role in quantum thermodynamics, where they serve as essential resources for optimizing work extraction, energy transfer, and the performance of microscopic thermal machines. In relativistic settings \cite{Fulling1973,Davies1975,Unruh1976}, the study of these correlations becomes particularly significant, as non-inertial motion can strongly influence the thermodynamic behavior of quantum systems. The UdW detector model \cite{Benatti2004,zhou2022steady,elghaayda2023entropy,koga2019entanglement}, which describes the interaction between point-like detectors and a quantum field, provides a powerful theoretical framework for investigating these phenomena. When a detector undergoes uniform acceleration, the Unruh effect causes it to perceive the quantum vacuum as a thermal bath, thereby introducing an intrinsic source of noise and decoherence. This perceived thermalization directly affects the quantum correlations shared between detectors and, consequently, modifies thermodynamic processes such as entropy production, heat flow, and work extraction. Thus, the use of the UdW model allows for a detailed exploration of how relativistic effects degrade, reshape, or even generate quantum correlations, offering new insights into quantum thermodynamics in accelerating frames and guiding the development of robust quantum technologies capable of operating in non-inertial environments.
	
	The primary objective of this work is to explore the hierarchy of quantum correlations-including quantum steering, entanglement, geometric quantum discord (GQD), and quantum coherence-within the framework of quantum thermodynamics for a two-UdW-detector system. We begin by analyzing the system in the absence and in the presence of decoherence, a crucial step for improving the robustness of qubits against environmental noise. Subsequently, by investigating the interplay between classical and quantum correlations in both Markovian and non-Markovian environments, we aim to enhance the stability of the system under various noise regimes. Furthermore, this study assesses the influence of variations in the energy level spacing of the atom ($\omega$) and the initial state selection parameter ($\Delta_0$) on the thermodynamic behavior of UdW devices, particularly as a function of temperature. Finally, we investigate this quantum system as a working substance in both a quantum heat engine and a refrigerator within the framework of the Stirling cycle.
	
	The structure of this article is organized as follows: 
	Section II (\hyperref[sec:model]{Model}) introduces the two-UdW-detector system and its thermal density operator, examining its dynamics under a correlated dephasing channel; 
	Section III (\hyperref[sec:quantum correlation]{Correlations}) provides a comprehensive overview of fundamental quantum concepts (steering, entanglement, geometric quantum discord, and coherence) and presents the detailed analysis and results of our study; 
	Section IV (\hyperref[sec:results]{Discussion}) discusses the implications of these findings within the broader research context; 
	Section V (\hyperref[sec:thermodynamics]{Thermodynamics}) addresses the Stirling cycle and evaluates its performance; 
	Section VI (\hyperref[sec:FEASIBILITY]{Feasibility}) discusses the experimental feasibility and possible physical implementations of the proposed model; 
	and finally, Section VII (\hyperref[sec:conclusion]{Conclusion}) concludes the article and provides additional discussion.

	\section{Theory and model} \label{sec:model}
	\subsection{Hamiltonian and matrix density}
	To explore the behavior of two uniformly accelerated UdW detectors in a 3 + 1-dimensional Minkowski spacetime \cite{Benatti2004,Bhuvaneswari2022,Elghaayda2025Quantum}, we describe the detectors using the framework of open quantum systems. In this approach, the system undergoes a non-unitary evolution of its density matrix due to decoherence and dissipation effects arising from its interaction with the surrounding environment. Each detector is represented as a two-level quantum system, and the overall dynamics are governed by the following total Hamiltonian \cite{Li2024,2025_Li} 
	\begin{equation}
		\mathcal{H} = \mathcal{H}_0 + \mathcal{H}_\phi + \mathcal{H}_I ,
	\end{equation}
	where, $\mathcal{H}_0$ represents the Hamiltonian of two independent detectors in the standard co-moving frame. The internal dynamics of each two-level detector is governed by a matrix $2\times 2$, and $\mathcal{H}_0$ can be concisely expressed as
	\begin{equation}
		\mathcal{H}_0=\frac{\omega}{2}\left(\sigma_3^{(A)} \otimes \mathbb{I}^{(B)}+\mathbb{I}^{(A)}\otimes\sigma_3^{(B)}   \right),
	\end{equation} 
	where $\omega$ denotes the energy level spacing of the atom and $\sigma_i^{(m)}$ being the Pauli matrices, and $m = A, B$ labels the different atoms. $\mathcal{H}_\phi$ is  the Hamiltonian of free massless scalar fields $\phi(t,x)$ satisfying the Klein-Gordon equation $\left( g^{\mu \nu }\nabla_\mu \nabla_\nu \right) \phi(t,x) = 0$. The interaction Hamiltonian can be written in dipole form \cite{Hu2013}
	\begin{equation}
		\mathcal{H}_I=\mu\left[  \left( \sigma_2^{(A)} \otimes \mathbb{I}^{(B)}\right)\phi(t,x_1) +\left( \mathbb{I}^{(A)}\otimes\sigma_2^{(B)}\right)\phi(t,x_2) \right], \label{3}
	\end{equation}
with $\mu$ is a small dimensionless coupling constant. 
	We study the time evolution of the detectors reduced density matrix, $\rho_{AB}(\tau)=\Tr_\phi(\rho_{tot})$, with 
	$\tau$ being the detectors proper time. The total system is initially separable, 
	$\rho_{tot}(0)=\rho_{AB}(0)\ket{0}\bra{0}$, where $\ket{0}$ denotes the field vacuum. As a closed system, its total density matrix follows the von Neumann equation $i\dot{\rho}_{AB}=\left[\mathcal{\mathcal{H}},\rho_{tot}(\tau) \right]$. In the weak-coupling limit, the open system dynamics of the detectors,  obey the Kossakowski–Lindblad master equation \cite{Gorini1976,Lindblad1976}
	\begin{equation}
		\frac{\partial\rho_{AB}(\tau)}{\partial \tau}=-i\left[\mathcal{H}_\text{eff},\rho_{AB}(\tau) \right]+\mathcal{L}\left[\rho_{AB}(\tau) \right],\label{4'}
	\end{equation}
	with 
	\begin{equation}
		\mathcal{H}_{\text{eff}}=\mathcal{H}_0-\frac{i}{2}\sum_{m=1}^{2}\sum_{n=1}^{2} C_{ij} \sigma_i^{(n)}\sigma_j^{(n)},
	\end{equation}
	and 
	\begin{equation}
	\mathcal{L}[\rho]=\sum_{i,j=1,2,3}^{}\sum_{\alpha,\beta=A,B}^{}\frac{C_{ij}}{2}\left[ 2\sigma_j^{(\beta)} \rho \sigma_i^{(\alpha)}-\{\sigma_i^{(\alpha)}\sigma_j^{(\beta)},\rho \} \right],
	\end{equation}
	to define the Kossakowski matrix $C_{ij}$, we first introduce the Wightman function of the scalar field, which is given by 
	\begin{equation}
		Y^+(x,x^\prime)=\bra{0}\phi (x)\phi(x^\prime)\ket{0}.
	\end{equation}
	The Fourier transform of the Wightman function is expressed as 
	\begin{equation}
		Y(\lambda)=\int_{-\infty}^{+\infty}d\tau \, e^{i\lambda \tau }Y^+(\tau )=\int_{-\infty}^{+\infty}d\tau \, e^{i\lambda \tau }\braket{\phi(\tau )}{\phi (0)}.\label{8}
	\end{equation}
	 The asymptotic equilibrium state of the two detectors, derived from the master equation (\ref{4'}), results from the competition between environmental dissipation in curved spacetime and quantum correlations generated through Markovian dynamics \cite{BenattiFloreanini2004PRA, BenattiFloreaniniPiani2003PRL}. For a bipartite system, the spatial separation $D = |x_{2} - x_{1}|$ acts as a control parameter in correlation generation, as the cross-Wightman functions—and thus the Kossakowski coefficients—depend on $D$ through $Y(\omega,D) = Y_0(\omega) f(\omega,D)$ \cite{Yu2011PRL, HuYu2013PRD}.
		Correlation generation is most effective for small separations and vanishes in the limit of infinite distance \cite{BenattiFloreanini2005JOB}, whereas a finite range of sufficiently small $D$ allows correlations to persist asymptotically, despite environmental dissipation \cite{BenattiFloreaniniMarzolino2010PRA}.
		Therefore, by restricting the analysis to small interatomic separations, the explicit distance dependence can be neglected; consequently, all Kossakowski matrices can be treated as equal, $C_{ij}^{(aa)} = C_{ij}^{(bb)} = C_{ij}^{(ab)} = C_{ij}^{(ba)} = C_{ij}$ \cite{HuYu2011}, where $\{a, b\}$ label the distinct detectors.%
		
	This Fourier transform allows for the determination of the coefficients $C_{ij}$ through the following decomposition
	
	\begin{equation}
		C_{ij}=\frac{\gamma_+}{2}\delta_{ij}-i\frac{\gamma_-}{2}\epsilon_{ijk} \delta_{3,k}+\gamma_0\delta_{3,i}\delta_{3,j},\label{cij}
	\end{equation}
	where  \begin{equation}
		\gamma_{\pm}=Y(\omega)\pm Y(-\omega), \,\,\, \gamma_0=Y(0)-\frac{\gamma_+}{2}\label{11}.
	\end{equation}
	Moreover, the interaction with the external scalar field
	induces a Lamb shift contribution to the effective Hamiltonian
	\begin{equation}
		H_{\text{eff}}=\frac{1}{2}\tilde{\omega}\sigma_3,
	\end{equation}		
	where $\tilde{\omega} = \omega + i\big[K(-\omega) - K(\omega)\big]$, and $K(\lambda)$ is the Hilbert
	transform of the Wightman function
	\begin{equation}
		K(\lambda) = \frac{1}{i\pi}\, \mathcal{P} \int_{-\infty}^{\infty}
		\frac{Y(\omega)}{\omega - \lambda}\, d\omega .
	\end{equation}
		
Along the trajectory of the accelerating detectors, the field’s Wightman function satisfies the Kubo-Martin-Schwinger (KMS) condition.
	\begin{equation}
		Y^{+}(\tau) = Y^{+}(\tau + i\beta),
	\end{equation}
where $\beta = 1/T$. The thermal nature of the detector response can be understood explicitly from the KMS condition. Specifically, it implies that the Fourier transform defined in Eq.~(\ref{8}) satisfies the relation
	\begin{equation}
		Y(-\omega) = e^{-\beta \omega} Y(\omega),
	\end{equation}
	which expresses the detailed-balance condition between the excitation and de-excitation processes of the two-level detector. Using Eq.~(\ref{11}), the coefficients $\gamma_{\pm} = Y(\omega) \pm Y(-\omega)$ consequently satisfy
	\begin{equation}
		\frac{Y(-\omega)}{Y(\omega)} = e^{-\beta \omega},
	\end{equation}
	leading directly to the ratio introduced in Eq.~(15),
	\begin{equation}
		\gamma = \frac{\gamma_-}{\gamma_+}
		= \tanh\left(\frac{\beta \omega}{2}\right).\label{16}
	\end{equation}
The detailed microscopic derivation of the Kossakowski coefficients $C_{ij}$ leading to Eq.~(\ref{cij}) is presented in Appendix~\ref{App:A}.
	
	This relation Eq.~(\ref{16}) is formally identical to the detailed balance condition for a two-level system in thermal equilibrium at temperature $T$. In the present context, $T$ corresponds to the Unruh temperature perceived by the uniformly accelerated detector. Therefore, although the scalar field is in the Minkowski vacuum state, the detector evolves toward a thermal stationary state determined by the parameter $\gamma$. In this sense, the two-level Unruh–DeWitt detector acts as an operational probe of the effective thermal nature of the vacuum in non-inertial frames.
	
	 In the Bloch representation of the UdW detector systems. Reconstructing the density matrix in the computational basis then gives the X-shaped stationary state  \cite{Bhuvaneswari2022}
	\begin{equation}
		\rho_{AB} =
		\begin{pmatrix}
			\rho_{11} & 0 & 0 & 0 \\
			0 & \rho_{22} & \rho_{23} & 0 \\
			0 & \rho_{32} & \rho_{33} & 0 \\
			0 & 0 & 0 & \rho_{44}
		\end{pmatrix},
	\end{equation}
	where 
	\begin{equation}
		\rho_{11}=\frac{\left(3+\Delta_0 \right)\left(\gamma-1 \right)^2  }{4\left(3+\gamma^2 \right) } \,, \rho_{22}=\rho_{33}=\frac{\left(3+\Delta_0 \right)\left(\gamma+1 \right)^2  }{4\left(3+\gamma^2 \right) } 
	\end{equation}
	\begin{equation}
		\rho_{44}=\frac{3-\Delta_0-\left(\Delta_0+1 \right)\gamma^2   }{4\left(3+\gamma^2 \right) }  \,,\,
		\rho_{23}=\rho_{32}=\frac{\Delta_0-\gamma^2 }{2\left(3+\gamma^2 \right) }.
	\end{equation}
	It is observed that the final equilibrium state of the two-detector system is determined by the ratio $\gamma$, which reflects the thermal effects associated with the Unruh phenomenon, as well as by the initial state parameters summarized by
	\[
	\Delta_0 = \sum_i \mathrm{Tr}\!\left[\rho_{AB}(0)\, \sigma_i^{A} \otimes \sigma_i^{B}\right].
	\]
	Requiring $\Delta_0$ to lie within the interval $-3 \leq \Delta_0 \leq 1$ ensures the non-negativity of the initial state $\rho_{AB}(0)$.
	
	\subsection{Preliminary Framework for Correlated Quantum Channels}
	
	This section formulates the evolution of a two-qubit thermal state under a correlated dephasing channel. The output state is obtained from the Kraus map \cite{NielsenChuang2010,HuFan2020,HuZhou2019} 
	\begin{equation}
		\rho_{AB}(t)=\sum_{i,j=0}^{3}L_{ij}\rho_{AB}(0)L_{ij}^\dagger \,\,,\,\, L_{ij}=\sqrt{p_{ij}}\left(\sigma_i\otimes \sigma_j \right),
	\end{equation}
	where  the joint probability is given by \cite{MacchiavelloPalma2002}
	\begin{equation}
		p_{ij}=\left(1-\mu  \right) p_ip_j+\mu p_i\delta_{ij},
	\end{equation}
	introduce the correlation parameter $\mu\in [0,1]$. For a pure dephasing process with $p_0=1-p$, $p_3=p$, the temporal behavior of coherence is governed by a random telegraph signal through \cite{HuZhou2019} 
	\begin{equation}
		p(t)=\frac{1-h(t)}{2},
	\end{equation}
	with
	\begin{equation}
		h(t)=	\begin{cases}
			e^{-\frac{t}{2\tau}}\left[ \cos\left(\frac{\nu t}{2\tau}\right)  +\frac{1 }{\nu} \sin\left(\frac{\nu t}{2\tau}\right) \right]  \,\,\,\,,\,\,\,4\tau > 1\\e^{-\frac{t}{2\tau}}\left[ \cosh\left(\frac{\nu t}{2\tau}\right)  +\frac{1 }{\nu} \sinh\left(\frac{\nu t}{2\tau}\right) \right]  \,\,\,\,,\,\,\,4\tau < 1,
		\end{cases}
	\end{equation}
	and $\nu =\sqrt{|4\tau^2-1|}$. The resulting attenuation of off-diagonal elements is described by
	\begin{equation}
		\zeta(t)=\left(1-\mu  \right) h^2(t)+\mu,
		\label{20}
	\end{equation}
	thus, the thermal density matrix evolves as
	\begin{equation}
		\rho_{AB}(t) =
		\begin{pmatrix}
			\rho_{11} & 0 & 0 & 0 \\
			0 & \rho_{22} & 	\zeta(t)\rho_{23} & 0 \\
			0 & 	\zeta(t)\rho_{32} & \rho_{33} & 0 \\
			0 & 0 & 0 & \rho_{44}
		\end{pmatrix}.
		\label{21}
	\end{equation}
	\section{Quantum resources measures}
	\label{sec:quantum correlation}
	\subsection{Quantum steering}
	Quantum steering represents a directional manifestation of quantum correlations, 
	whereby local measurements on one qubit can non-locally influence the state of the other. Following the entropic formulation proposed by \textit{Schneeloch et al.}~\cite{Schneeloch2013}, the steering criterion is expressed as
	\begin{equation}
		f_{AB} = H_x(B|A) + H_y(B|A) + H_z(B|A) \ge 2,
		\label{eq:steering_inequality}
	\end{equation}
	where $H_i(B|A)$ denotes the conditional Shannon entropy corresponding to 
	Pauli-basis measurements. 
	A violation of this bound ($f_{AB} < 2$) indicates the existence of steerability 
	between the two subsystems.
	
	The quantitative measure of steering is given by \cite{Jaloum2025WorkExtraction,Jaloum2025}
	\begin{equation}
		\label{23}
		S_{A \rightarrow B} = \max\!\left(0, \frac{  f_{AB}-2}{f_{\max}-2}\right), 
		\qquad f_{\max} = 6,
	\end{equation}
	where \begin{align}
		f_{AB}=&\frac{1}{2}\sum_{i=1}^{4}\{f_{x_{i}}^{AB}log_{2}(f_{x_{i}}^{AB})+f_{y_{i}}^{AB}log_{2}(f_{y_{i}}^{AB})+f_{z_{i}}^{AB}log_{2}(f_{z_{i}}^{AB})\}\nonumber\\
		-&\sum_{i=1}^{2}\{f_{x_{i}}^{A}log_{2}(f_{x_{i}}^{A})+f_{y_{i}}^{A}log_{2}(f_{y_{i}}^{A})+f_{z_{i}}^{A}log_{2}(f_{z_{i}}^{A})\},
	\end{align}
	and
	\begin{align*}
		&f_{x_{1}}^{AB}=f_{x_{2}}^{AB}=1+2{\rho}_{23},\,\,\,\,\,
		f_{x_{3}}^{AB}=f_{x_{4}}^{AB}=1-2{\rho}_{23},\\
		&f_{y_{1}}^{AB}=f_{y_{2}}^{AB}=1+2{\rho}_{23},\,\,\,\,\,
		f_{y_{3}}^{AB}=f_{y_{4}}^{AB}=1-2{\rho}_{23},
		\\
		&f_{z_{i}}^{AB}=4{\rho}_{i,i}\,,\,\,\,\, f_{x_{1}}^{A}=f_{x_{2}}^{A}=1 ,\,\,\,f_{y_{1}}^{A}=f_{y_{2}}^{A}=1, \\
		&{f}_{z{_1}}^{A}={f}_{z_{2}}^{A}=1\pm({\rho}_{11}-{\rho}_{44}),
	\end{align*}
	and the asymmetry between both directions is characterized by
	\begin{equation}
		\Delta_{12} = \left| S_{A \rightarrow B} - S_{B \rightarrow A} \right|.
		\label{eq 25}
	\end{equation}
	For the considered UdW system, the structure of the Hamiltonian guarantees that $S_{A \rightarrow B} = S_{B \rightarrow A}$, indicating a bidirectional (two-way) steering. This formalism provides a rigorous framework for quantifying directional quantum correlations and for establishing the hierarchy among steering, entanglement, and other coherence-based measures in systems~\cite{Schneeloch2013}.
	
	\subsection{Entanglement of formation}
	
	The entanglement of formation, \(\xi\), for a bipartite pure state is defined as \cite{Hill1997}
	\begin{equation}
		\xi({\rho_{AB}}) = \mathcal{E} \left(C({\rho_{AB}})\right),
	\end{equation}
	with
	\begin{equation}
		\mathcal{E} \left(C({\rho_{AB}})\right)
		= L \left[
		\frac{1+\sqrt{1-C^{2}({\rho_{AB}})}}{2}
		\right],
	\end{equation}
	where $L(x) = -x\log_{2}x - (1-x)\log_{2}(1-x)$
	and \(C\) denotes the concurrence. To measure the concurrence, we employ Wootters concurrence formalism~\cite{Wootters1998,HillWootters1997,Adesso2004,Vidal2002,Plenio2005}. 
	For a general two-qubit density matrix ${\rho}_{AB}$, the concurrence is defined as
	\begin{equation}
		C({\rho}_{AB}) = \max \left\{ 0, \sqrt{\nu_1} - \sqrt{\nu_2} - \sqrt{\nu_3} - \sqrt{\nu_4} \right\},
		\label{26}
	\end{equation}
	where $\nu_i$ $(i = 1, \dots, 4)$ are the eigenvalues, in decreasing order, of the non-Hermitian matrix 
	\begin{equation}
		{R} = {\rho}_{AB} \, (\sigma_y \otimes \sigma_y) \, {\rho}_{AB}^{*} \, (\sigma_y \otimes \sigma_y),
	\end{equation}
	with ${\rho}_{AB}^{*}$ being the complex conjugate of ${\rho}_{AB}$ in the computational basis 
	and $\sigma_y$ the Pauli matrix.
	For the considered $X$-type state, the concurrence can be expressed analytically as
	\begin{equation}
		C({\rho}_{AB}) = 2 \max \left\{ 0, \, |\rho_{23}| - \sqrt{\rho_{11}\rho_{44}}, \,  - \sqrt{\rho_{22}\rho_{33}} \right\},
	\end{equation}
	where $\rho_{ij}$ denote the matrix elements of the density operator ${\rho}_{AB}$. The concurrence ranges from $C = 0$ ($\xi=0$), corresponding to a separable state, to $C = 1$ $(\xi=1)$, representing a maximally entangled (Bell) state. This measure provides a reliable quantification of bipartite entanglement, particularly suitable for $X$-structured density matrices arising in UdW systems.

	\subsection{Geometric quantum discord}
	
	The geometric approach constitutes a powerful framework for quantifying quantum correlations even beyond entanglement. In this context, the \textit{geometric quantum discord} (GQD) is defined as the minimal Schatten 1-norm distance \cite{PaulaOliveiraSarandy2013} between a bipartite quantum state ${\rho}_{AB}$ and the closest classical–quantum state ${\rho}_c$ \cite{NakanoPianiAdesso2013,CiccarelloTufarelliGiovannetti2014}
	\begin{equation}
		Q_G({\rho}_{AB}) = \min_{{\rho}_c \in \Omega} \| {\rho}_{AB} - {\rho}_c \|_1 ,
		\label{eq:GQD_def}
	\end{equation}
	where $\|X\|_1 = \mathrm{Tr}[\sqrt{X^\dagger X}]$ denotes the trace norm, and $\Omega$ denotes the set of all classical-quantum states, each of which can be represented in the form
	\begin{equation}
		{\rho}_c = \sum_k p_k \, \Pi_{k,1} \otimes {\rho}_{k,2},
		\label{eq:classical_state}
	\end{equation}
	where $\{p_k\}$ is a probability distribution such that $\sum_k p_k = 1$, $\Pi_{k,1}$ are the orthogonal projectors associated with qubit 1, and ${\rho}_{k,2}$ are the density operators of qubit 2. For a general two-qubit $X$ state, the geometric quantum discord based on the Schatten 1-norm can be writes as \cite{JaloumAmazioug2025BES,AmaziougDaoud2024}
	
	\begin{equation}
		Q_G=\frac{1}{2} \sqrt{\frac{F_{11}^2\max\{F_{22}^2+F_{30}^2,F_{33}^2\}-F_{22}^2\min\{F_{11}^2,F_{33}^2\}}{\max\{F_{22}^2+F_{30}^2,F_{33}^2\}-\min\{F_{11}^2,F_{33}^2\}+F_{11}^2-F_{22}^2}}.
		\label{31}
	\end{equation}
	The {Fano-Bloch} representation $F$ of the two-qubit state provides the correlation matrix components $F_{\mu\nu}$ as
	\begin{equation}
		F= \frac{1}{4} \,\sum_{\mu ,\nu=0}^{3} F_{\mu\nu}  \, (\sigma_\mu \otimes \sigma_\nu),
		\label{eq:R_components}
	\end{equation}
	where $\sigma_\mu$ ($\mu=0,1,2,3$) are the Pauli matrices, with $\sigma_0$ being the identity operator. For the considered thermal $X$-state of the two  qubits, the non-vanishing elements of $F_{\mu\nu}$ are given by
	\begin{align}
		F_{11} &= 2 \rho_{23}, \quad
		F_{22} = 2 \rho_{23}, \\
		F_{33} &= 1 -2\left(  \rho_{22} + \rho_{33} \right) , \quad
		F_{30} = 	2\left( \rho_{11} + \rho_{22}\right)  - 1.\nonumber
	\end{align}
	\subsection{Quantum Coherence}
	
	Quantum coherence represents a fundamental resource in quantum information theory, reflecting the superposition principle that underpins all quantum phenomena. Following Baumgratz \textit{et al.}~\cite{BaumgratzCramerPlenio2014}, it can be quantified by the $l_{1}$-norm of coherence, which measures the minimal distance between a given quantum state and the set of incoherent states. For a bipartite density matrix ${\rho}_{AB}$ expressed in the computational basis as 
	\begin{align}
		{\rho}_{AB} = \sum_{i,j} \rho_{ij} \, |i\rangle \langle j| ,
	\end{align}
	the $l_{1}$-norm of coherence is defined by the sum of the absolute values of the off-diagonal elements
$
		C_{l_{1}}({\rho}_{AB}) = \sum_{i \neq j} |\rho_{ij}| .
		\label{eq:l1norm}
$

	In the case of the UdW system considered here, the $l_{1}$-norm coherence can be explicitly evaluated from the density matrix elements as
	\begin{align}
		C_{l_{1}}({\rho}_{AB}) = 2 |\rho_{23}|,
		\label{36}
	\end{align}
	this expression quantifies the degree of quantum coherence present in the thermal state of the coupled  qubits. A higher value of $C_{l_{1}}$ indicates a stronger superposition between the basis states, reflecting enhanced capability for quantum interference and information processing.
	\vspace*{-0.4cm}
	\section{Results and discussions}
	\label{sec:results}
	\subsection{Steady state}
	In the steady-state regime, the steerability between qubits $A$ (Alice) and $B$ (Bob) can manifest in different configurations depending on the sign of $\Delta_{12}$.For $\Delta_{12} > 0$, the system exhibits one-way steering, meaning either $S_{A \rightarrow B} > 0$ and $S_{B \rightarrow A} = 0$ (Alice can steer Bob but not vice versa), or $S_{A \rightarrow B} = 0$ and $S_{B \rightarrow A} > 0$ (the opposite situation).When $\Delta_{12} = 0$, the system allows either two-way steering ($S_{A \rightarrow B} = S_{B \rightarrow A} > 0$) or no-way steering ($S_{A \rightarrow B} = S_{B \rightarrow A} = 0$), the latter occurring even when the qubits remain entangled.In our model, the equalities in Equations (\ref{23}) and (\ref{eq 25}) yield $\Delta_{12}=0$, indicating a bidirectional steering symmetry between qubits $A$ and $B$ (see Figure \ref{fig:steering}(c)).
	\onecolumngrid
	\begin{center}
		\vspace*{0.4cm}
		\begin{minipage}{0.32\textwidth}
			\centering
			\includegraphics[width=\linewidth]{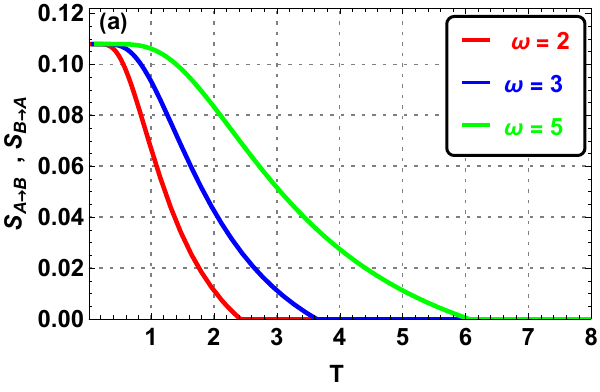}
		\end{minipage}
		\begin{minipage}{0.32\textwidth}
			\centering
			\includegraphics[width=\linewidth]{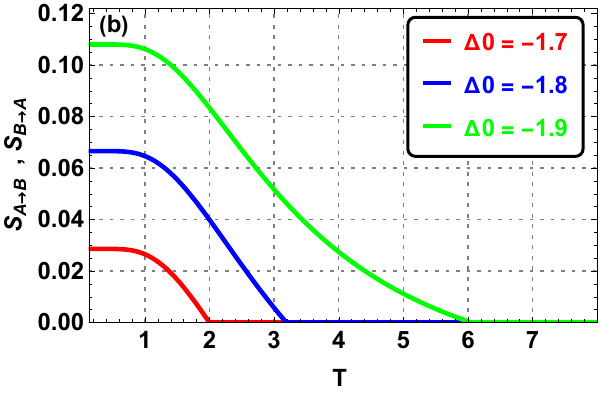} 
		\end{minipage}
		\begin{minipage}{0.32\textwidth}
			\centering
			\includegraphics[width=\linewidth]{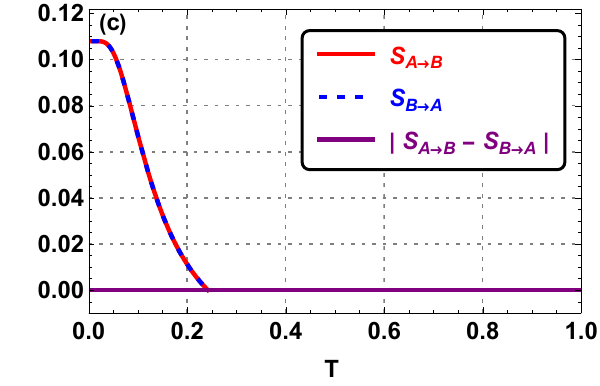}
		\end{minipage}
		\captionof{figure}{(a) Quantum steerability \(S_{A \rightarrow B}\) and \(S_{B \rightarrow A}\) as a function of temperature \(T\)  for different values of  $\omega$ with fixed \(\Delta_0 = -1.9\); 
			(b) Quantum steerability versus \(T\)  for different values of the initial state
			parameter $\Delta_0$ with $\omega = 5$; 
			(c) Comparison of $S_{A \rightarrow B}$, $S_{B \rightarrow A}$, and $\Delta_{12}$ for $\Delta_0 =-1.9 $ and $\omega =0.2$.}
		\label{fig:steering}
	\end{center}
	\newpage
	\twocolumngrid

	In Fig.~\ref{fig:steering}(a), we present the quantum steerabilities $S_{A \rightarrow B}$ and $S_{B \rightarrow A}$ as functions of the temperature $T$ for different values of  $\omega$, while keeping the initial state parameter \(\Delta_0=-1.9\). Owing to the symmetry of the system, both steerabilities coincide, i.e., $S_{A \rightarrow B} = S_{B \rightarrow A}$. For temperatures $T \in [0, 0.4]$, the steerability remains maximal for all values of $\omega$. Furthermore, the threshold temperature at which the steerability begins to decline shifts to higher values as $\omega$ decreases. In the range $T \in [0.4, 6]$, the steerability exhibits a monotonic decrease, while its magnitude becomes larger for smaller values of $\omega$, as illustrated in Fig.~\ref{fig:steering}(a). Additionally, for $T > 6$, the steerability vanishes, which can be attributed to decoherence effects~\cite{Zurek2003}. 
	
	The quantum steerabilities $S_{A \rightarrow B}$ and $S_{B \rightarrow A}$ are investigated as functions of the temperature $T$ for different values of the initial state parameter $\Delta_0$, while the energy is kept fixed at $\omega = 5$ (Fig.~\ref{fig:steering}(b)). The results indicate that quantum steerability decays rapidly in the presence of thermal noise. We also observe that the threshold temperature at which steerability vanishes
	\onecolumngrid
	\begin{center}
		\vspace*{0.4cm}
		\begin{minipage}{0.32\textwidth}
			\centering
			\includegraphics[width=\linewidth]{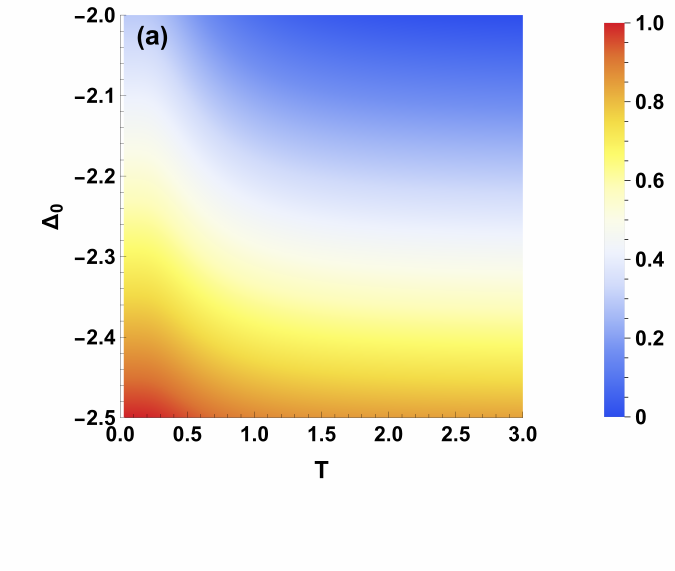}
		\end{minipage}
		\begin{minipage}{0.32\textwidth}
			\centering
			\includegraphics[width=\linewidth]{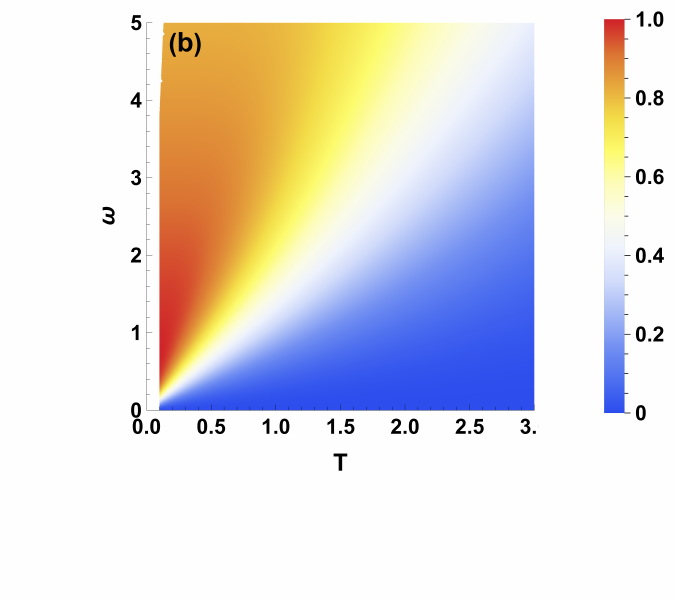} 
		\end{minipage}
		\begin{minipage}{0.32\textwidth}
			\centering
			\includegraphics[width=\linewidth]{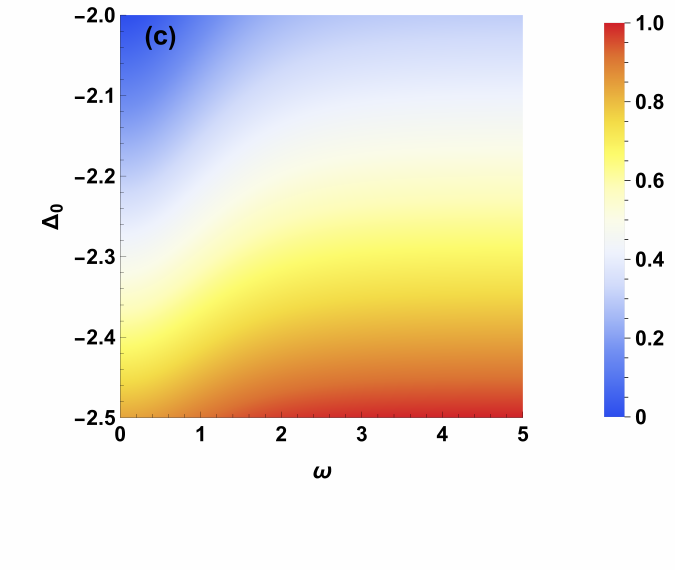}
		\end{minipage}
	\vspace*{-1.1cm}	
		\captionof{figure}{(a) Quantum steering  versus $\Delta_0$ and $T$ with $\omega=1$; (b) Quantum steering  versus $\omega$ and $T$ with $\Delta_0=-2.2$; (c) Quantum steering  versus $\Delta_0$ and $\omega$ with $T=0.8$.	\label{2}}
	\end{center}
	\twocolumngrid
	increases with increasing $\Delta_0$. At low 
	temperatures, the steerability becomes larger for smaller values of $\Delta_0$. 
	Nevertheless, higher the energy level spacing of the atom allow the steerability to persist over a broader temperature range.
	
	In Fig.~\ref{fig:steering}(c), we plot the steerabilities $S_{A \rightarrow B}$, $S_{B \rightarrow A}$, along with the asymmetry $\Delta_{12}$ as functions of the temperature $T$. We observe that $S_{A \rightarrow B} = S_{B \rightarrow A} > 0$ (i.e., $\Delta_{12} = 0$) for $T \in [0, 0.25]$. This indicates the presence of two-way quantum steering between qubit $A$ and qubit $B$.  Moreover, for $T \in [0.25, 1]$, we have {$S_{A \rightarrow B} = S_{B \rightarrow A} =0$}, effectively indicating negligible steering between the two qubits. 
	
	The three density plots in Fig.~\ref{2} illustrate how the quantum steering measure depends on the temperature $T$ as well as on the parameters $\Delta_{0}$ and $\omega$.
	The results show that quantum steering persists only within a limited low-temperature 
	region, where quantum correlations remain sufficiently strong. {As the temperature increases, thermal noise becomes progressively more dominant, acting as a strong source of decoherence that disrupts the  quantum correlations. Consequently, the steering measure decreases continuously, since higher temperatures weaken the nonclassical features required to sustain directional correlations. When the thermal fluctuations become sufficiently strong, these correlations are effectively washed out, and quantum steering ceases to be observable in the high-temperature regime.}
	The parameters $\Delta_{0}$  and $\omega $ {play} a central role in determining both 
	the magnitude and the robustness of steering, as its variation modifies the structure 
	of the density matrix and controls the extent of the region where steering survives. 
	Overall, the three plots highlight the strong sensitivity of quantum steering to 
	thermal fluctuations and the crucial role of $\Delta_{0}$ and $\omega$ in sustaining directional 
	quantum correlations.
	\onecolumngrid
	\begin{center}
		\vspace*{0.4cm}
		\begin{minipage}{0.48\textwidth}
			\centering
			\includegraphics[width=\linewidth]{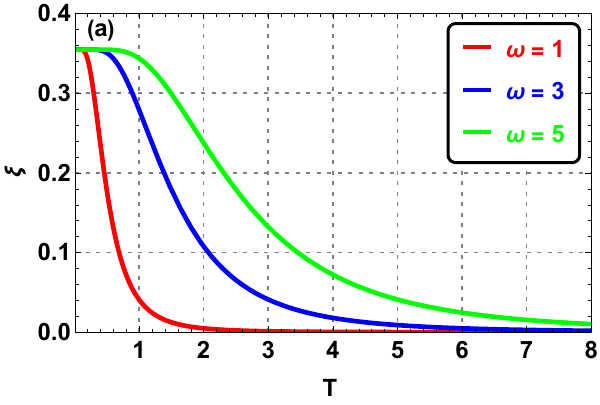}
		\end{minipage}
		\begin{minipage}{0.48\textwidth}
			\centering
			\includegraphics[width=\linewidth]{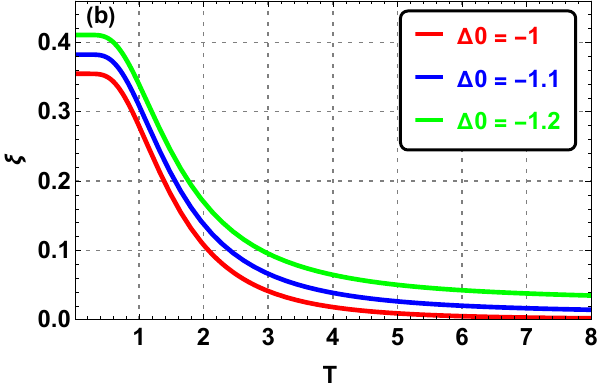} 
		\end{minipage}\\
		\captionof{figure}{ (a) Entanglement of formation  \(\xi\)  versus temperature \(T\) for different values of \(\omega\) with  \(\Delta_0 = -1\); 
			(b)  entanglement of formation  versus \(T\) for different values of \(\Delta_0\) with \(\omega= 3\).	\label{Fig3}}

	\end{center}
	\newpage
	\twocolumngrid
	The behavior of the entanglement of formation $\xi$ as a function of temperature $T$ for several values of $\omega$ is examined in Fig.~\ref{Fig3}(a), with the initial state parameter fixed at $\Delta_0=-1$. {The results demonstrate that, as $T \to 0$, the entanglement reaches its maximum and becomes effectively independent of temperature. As expected, entanglement decreases monotonically with rising temperature due to thermal noise. Moreover,  larger energy spacing $\omega$ helps sustain higher entanglement values at finite temperatures, demonstrating that entanglement is more robust against thermal effects as $\omega$ increases. These observations are consistent with the experimental findings on thermal entanglement reported in Refs.~\cite{Shaw2009,Tian2011}.}
	
	In Fig.~\ref{Fig3}(b), we present the temperature dependence of the entanglement of formation $\xi$ for different values of  initial state 	parameter $\Delta_0$,  Furthermore, the entanglement rapidly diminishes and tends to zero for sufficiently high values of $T$. Interestingly, in the intermediate temperature interval $T \in [0, 0.4]$, the entanglement increases as $\Delta_0$ decreases, as illustrated in Fig.~\ref{Fig3}(b). For {temperatures $T > 5$}, the entanglement  vanishes for different values of different values of $\Delta_0$.
	\onecolumngrid
	\begin{center}
		\vspace*{0.2cm}
		\begin{minipage}{0.48\textwidth}
			\centering
			\includegraphics[width=\linewidth]{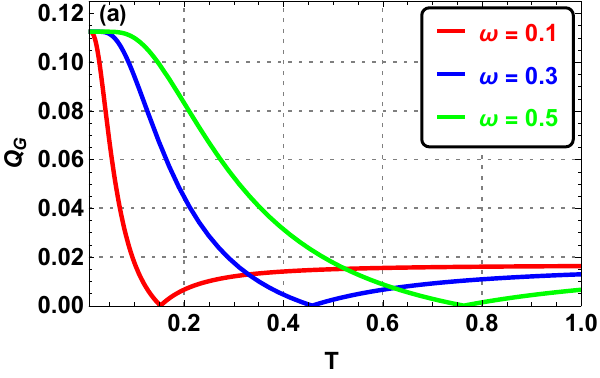}
		\end{minipage}
		\begin{minipage}{0.48\textwidth}
			\centering
			\includegraphics[width=\linewidth]{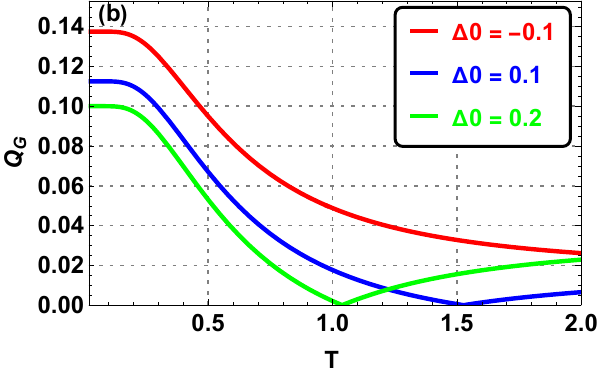} 
		\end{minipage}
		\vspace*{-0.2cm}
		\captionof{figure}{Geometric quantum discord $Q_G$ as a function of the temperature $T$; 
			(a) for different values of $\omega$ with $\Delta_0 = 0.1$. 
			(b) for different values of $\Delta_0$ with $\omega = 1$.	\label{fig:4}}
	\end{center}
	\twocolumngrid

	In Fig.~\ref{fig:4}(a), The results show that the geometric quantum discord $Q_G$ exhibits a non-monotonic behavior with temperature $T$: it initially decreases due to thermal fluctuations, then increases at higher temperatures. This behavior suggests that, despite the initial loss of correlations at low temperatures, certain thermal effects promote a partial restoration of quantum correlations at intermediate temperatures. The parameter $\Delta_0$ sets the baseline correlation level, while the energy spacing $\omega$ modulates the depth and position of the minimum in $Q_G$: larger values of $\omega$ tend to mitigate the initial decay and enhance the persistence of quantum correlations.
	
	The results in Fig.~\ref{fig:4}(b) indicate that the geometric quantum discord $Q_G$ is highly sensitive to the initial state parameter $\Delta_0$. For all considered values of $\Delta_0$, $Q_G$ exhibits a non-monotonic behavior with temperature $T$, decreasing at low temperatures due to thermal fluctuations and then increasing at higher temperatures. The magnitude and position of the minimum of $Q_G$ are strongly influenced by $\Delta_0$: larger values of $\Delta_0$ lead to a higher baseline discord and a shallower initial decay, indicating that the choice of the initial state can enhance the robustness of quantum correlations against thermal effects. These observations emphasize the crucial role of the initial state configuration in shaping the thermal dynamics of quantum discord.
	\onecolumngrid
	\begin{center}
		\vspace*{0.4cm}
		\begin{minipage}{0.48\textwidth}
			\centering
			\includegraphics[width=\linewidth]{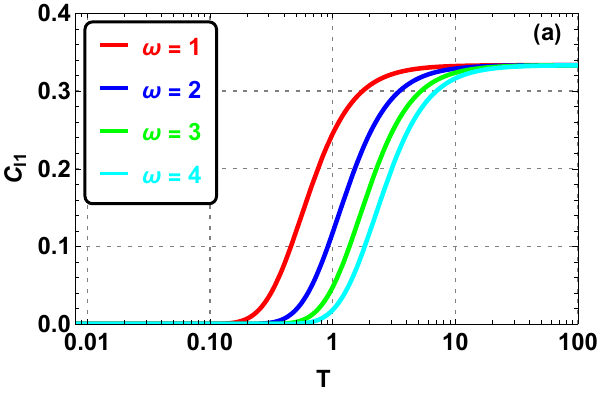}
		\end{minipage}
		\begin{minipage}{0.48\textwidth}
			\centering
			\includegraphics[width=\linewidth]{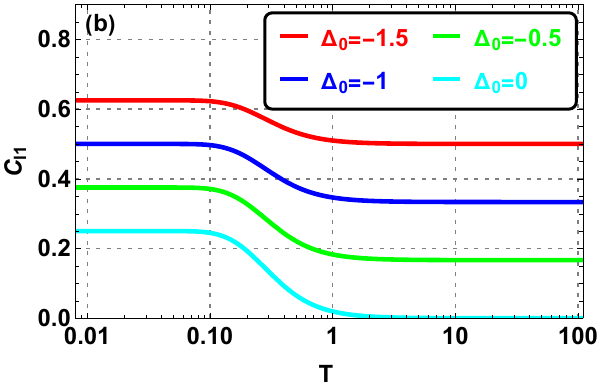} 
		\end{minipage}
		\captionof{figure}{The quantum coherence $C_{l1}$ as a function of temperature $T$ is presented: 
			(a) for different values of the  energy $\omega$, with  initial state
			selection parameter \(\Delta_0=1\) ;
			(b)for different values of  the initial state
			selection parameter \(\Delta_0\), with $\omega=0.5$.}
		\label{fig:5}
	\end{center}
	\newpage
	\twocolumngrid
The results in the figure~\ref{fig:5}(a) show that the quantum coherence $C_{l_1}$ increases with temperature $T$, approaching an asymptotic maximum value for all considered values of $\omega$. At low temperatures, the coherence is initially small, indicating limited quantum correlations. The energy spacing $\omega$ affects the rate at which $C_{l_1}$ reaches its maximum: smaller values of $\omega$ allow the coherence to saturate at lower temperatures, whereas larger $\omega$ delay this saturation. These observations suggest that the energy separation plays a crucial role in determining the temperature scale over which quantum coherence.
	
	Figure~\ref{fig:5}(b) illustrates the quantum coherence $C_{l1}$ as a function of temperature $T$ for various values of the parameter $\Delta_0$. 
		The coherence generally decreases with increasing temperature, highlighting the suppressive effect of thermal fluctuations on quantum superpositions. 
		At low temperatures ($T \lesssim 0.1$), $C_{l1}$ reaches its maximum values, indicating strong coherent superpositions between the qubit states. 
		As the temperature increases, the coherence gradually diminishes, approaching lower values at high temperatures ($T \sim 100$), where thermal noise dominates and superpositions are largely destroyed. Comparing different $\Delta_0$ values, larger (less negative) interaction parameters 	maintain higher coherence at low temperatures. In particular,
		the curve for $\Delta_0 = -1.5$ exhibits the highest initial coherence, while more negative values ($\Delta_0 =-1, -0.5, 0$) show progressively smaller coherence. 
		This indicates that the interaction strength significantly affects the robustness of 	quantum coherence against thermal effects, emphasizing the role of tuning $\Delta_0$ to preserve quantum superpositions in practical quantum systems.
	\onecolumngrid
	\begin{center}
			\centering
			\includegraphics[scale=0.58]{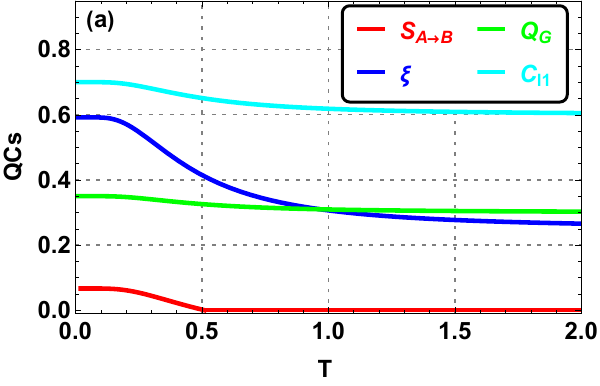}
			\includegraphics[scale=0.58]{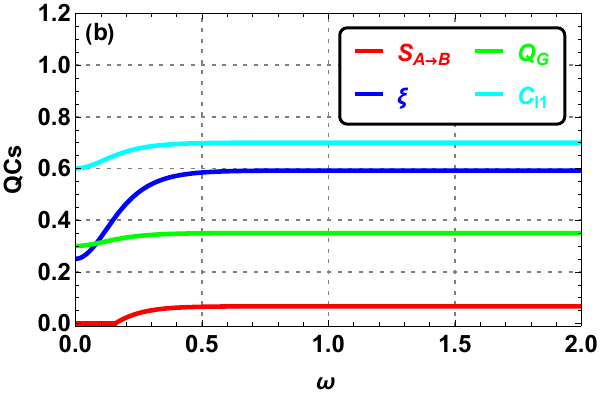}
			\includegraphics[scale=0.58]{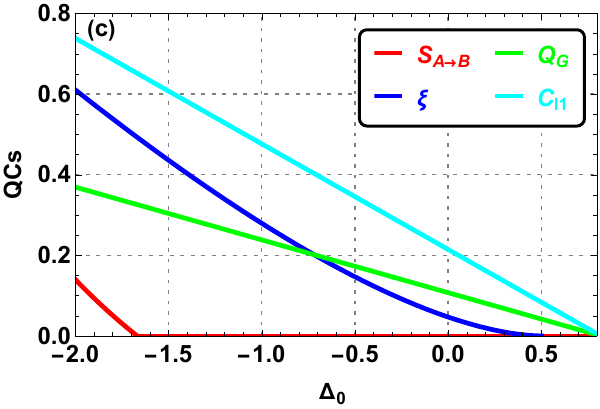} 
		\captionof{figure}{Plot of quantum coherence ($C_{l1}$) and quantum correlations, including quantum steering ($S_{A \to B}$), entanglement of formation ($\xi$), geometric quantum discord ($Q_G$): 
			(a) as a function of temperature ($T$) with $\Delta_0= -1.8$ and $\omega= 0.8$; 
			(b) as a function of the  energy ($\omega$) at $T = 0.1$ with $\Delta_0= -1.8$; 
			and (c) as a function of  ($\Delta_0$) at $T = 0.5$ with $\omega = 1.5$.
		\label{fig:6}}
	\end{center}
	\twocolumngrid

Figures~\ref{fig:6}(a) and~\ref{fig:6}(b) jointly illustrate the influence of temperature and energy spacing on quantum coherence and quantum correlations in the system. As shown in Fig.~\ref{fig:6}(a), at low temperatures all quantifiers reach high values, revealing strong entanglement and robust nonclassical correlations. With increasing temperature, thermal fluctuations progressively degrade these resources, in accordance with the hierarchy of quantum correlations: quantum steering is the most fragile and disappears at moderate temperatures, whereas entanglement and quantum discord persist over a wider thermal range, and quantum coherence remains finite throughout. In contrast, Fig.~\ref{fig:6}(b) demonstrates that increasing the energy spacing parameter $\omega$ enhances all quantum correlations, which rise rapidly for small $\omega$ and then saturate beyond a critical value $\omega \simeq 0.5$. This saturation marks an optimal regime where quantum resources are maximized and become insensitive to further increases in $\omega$, highlighting the complementary roles of temperature and spectral tuning in controlling the robustness and optimization of quantum correlations.

	As shown in Fig.~\ref{fig:6}(c), quantum coherence and all four quantum correlation measures decrease with increasing $\Delta_0$. While $\xi$ and $C_{I_1}$ remain relatively robust, steering ($S_{A \to B}$) vanishes earlier, around $\Delta_0 \approx -1.5$, highlighting its fragility. All measures reach zero near $\Delta_0 \approx 0.6$, marking the transition to a classical state. This behavior identifies $\Delta_0$ as an intrinsic decoherence parameter driving the progressive classicalization of the system.  We remark also, that  increasing the parameter $\Delta_0$ gradually suppresses the more ``fragile'' and resourceful quantum correlations, such as steering (which is linked to nonlocality), while leaving the more fundamental and robust quantum features, like discord and coherence, largely unaffected, as these do not require strong nonlocal correlations.
	\subsection{Dynamical state}
	In this section, we examine the dynamics of quantum resources in a UdW system subjected to correlated dephasing channels. 
	\onecolumngrid 
	\begin{center}
		\vspace*{0.08cm}
		\begin{minipage}{0.48\textwidth}
			\centering
			\includegraphics[width=\linewidth]{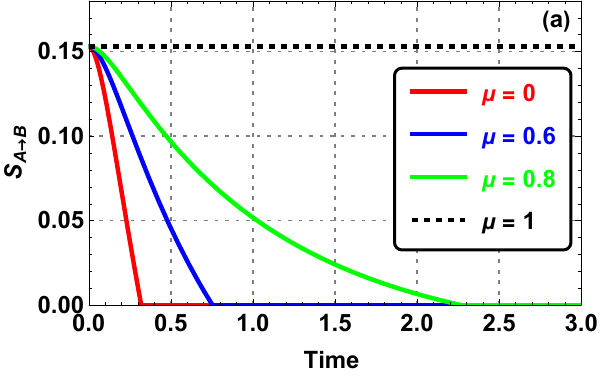}
		\end{minipage}
		\begin{minipage}{0.48\textwidth}
			\centering
			\includegraphics[width=\linewidth]{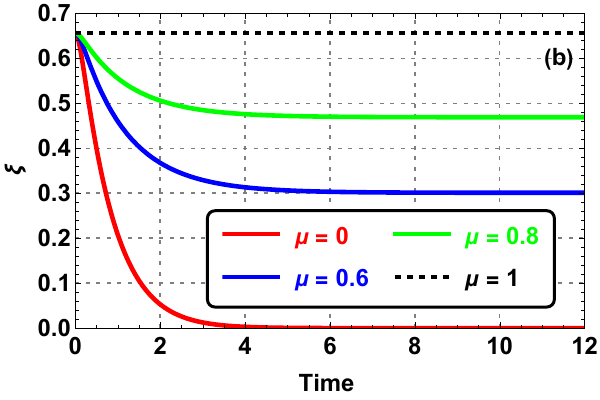}
		\end{minipage}
		\\
		\begin{minipage}{0.48\textwidth}
			\centering
			\includegraphics[width=\linewidth]{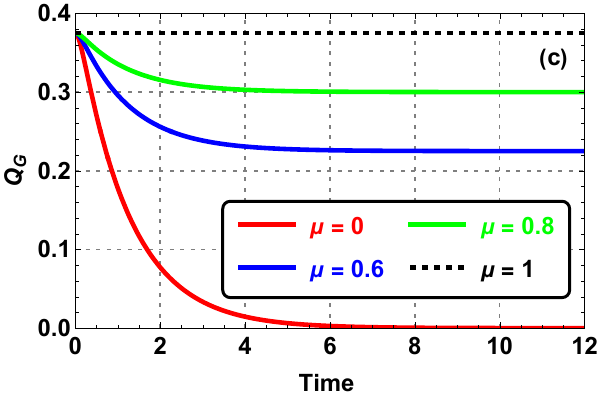}
		\end{minipage}
		\begin{minipage}{0.48\textwidth}
			\centering
			\includegraphics[width=\linewidth]{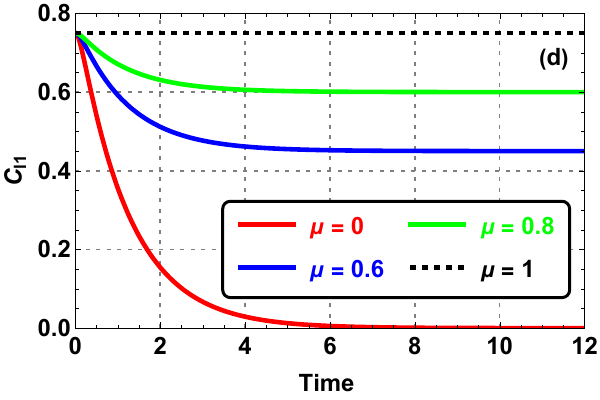}
		\end{minipage}
		\captionof{figure}{Dynamical evolution of quantum steering $S_{A\to B}$ (a), entanglement of formation $\xi$ (b), geometric quantum discord $Q_G$ (c), and quantum coherence $C_{l1}$ (d) in the Markovian regime with $\tau = 0.1$, $T = 0.1$, $\Delta_0 = -2$, and $\omega = 1$. For all curves, the red, blue, green, and black curves correspond to $\mu = 0$, $0.6$, $0.8$, and $1$, respectively.	\label{FiG 7}}
	\end{center}
	\twocolumngrid
We provide a detailed analysis of the influence of classical correlations, quantified by the parameter $\mu$, on the time evolution of quantum steering, entanglement, geometric quantum discord, and quantum coherence, in both Markovian and	 non-Markovian regimes. Our results reveal a significant enhancement of these quantum resources as the degree of classical correlations increases.
	 
	Starting from the thermal time-dependent state given in Equation (\ref{21}), analytical expressions for quantum steering, entanglement, geometric quantum discord, and quantum coherence are obtained by substituting ${\rho}_{23}$ with $\zeta{\rho}_{23}$, respectively, in Equations (\ref{23}), (\ref{26}), (\ref{31}), and (\ref{36}), where $\zeta(t)=\left(1-\mu \right) h^2(t)+\mu$. The results displayed in Figure \ref{FiG 7} demonstrate the crucial role played by classical correlations in enhancing quantum resources within a Markovian environment. Increasing $\mu$ yields a marked improvement in quantum steering, entanglement, geometric quantum discord, and quantum coherence. For $\mu<1$, these quantities decay monotonically and exponentially over time, indicating that weaker classical correlations amplify decoherence effects. In particular, when $\mu=0$, Equation (\ref{20}) reduces to $\zeta(t)=h^{2}(t)$, and all quantum resources rapidly decay toward their steady-state values. In contrast, for a fully correlated dephasing channel ($\mu=1$), all quantifiers remain constant, reflecting complete robustness against decoherence. These findings highlight the importance of classical correlations in preserving quantum coherence and improving the performance of quantum systems in information-processing tasks.
	
	\onecolumngrid
	\begin{center}
		\begin{minipage}{0.48\textwidth}
			\centering
			\includegraphics[width=\linewidth]{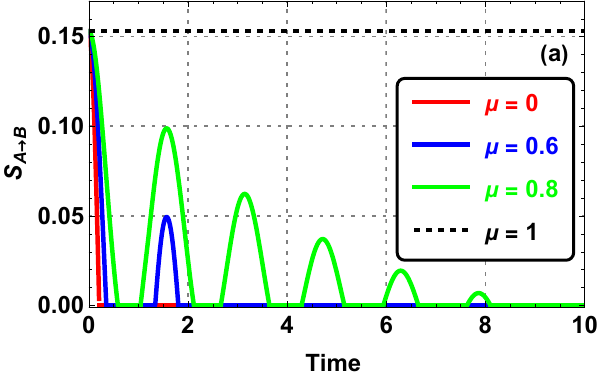}
		\end{minipage}
		\begin{minipage}{0.48\textwidth}
			\centering
			\includegraphics[width=\linewidth]{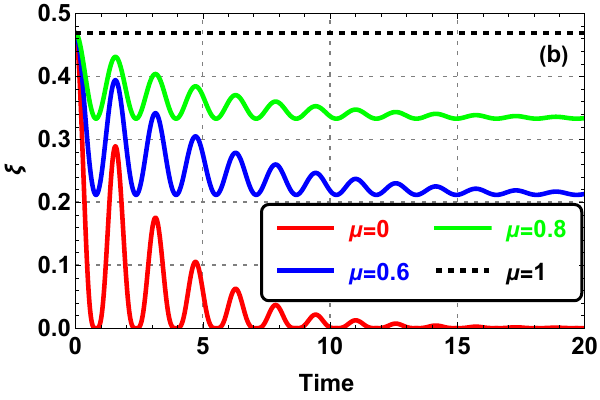}
		\end{minipage}
		\\
		\begin{minipage}{0.48\textwidth}
			\centering
			\includegraphics[width=\linewidth]{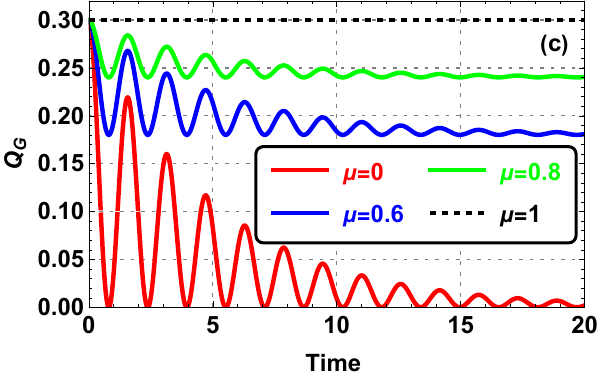}
		\end{minipage}
		\begin{minipage}{0.48\textwidth}
			\centering
			\includegraphics[width=\linewidth]{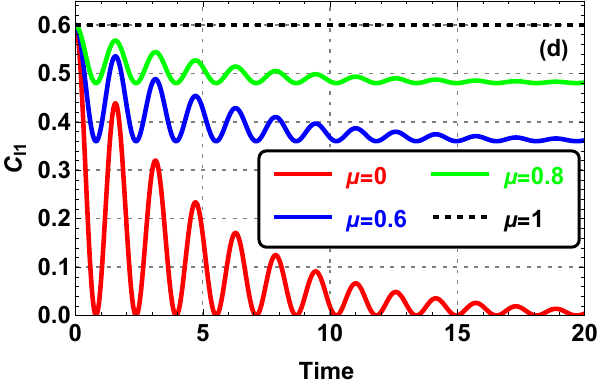}
		\end{minipage}
		\captionof{figure}{Time evolution of quantum steering $S_{A\to B}$ (a), entanglement of formation $\xi$ (b), geometric quantum discord $Q_G$ (c), and quantum coherence $C_{l1}$ (d) in the Non-Markovian regime with $\tau =5$, $T = 0.1$, $\Delta_0 = -2$, and $\omega = 1$. For all curves, the red, blue, green, and black curves correspond to $\mu = 0$, $0.6$, $0.8$, and $1$, respectively.
		}
		\label{Fig 8}
	\end{center}
	\twocolumngrid

	Figure \ref{Fig 8} extends the analysis to the non-Markovian regime.  This figure shows that, for all values $\mu<1$, the quantum resources quickly decay to zero, whereas for $\mu=1$, they remain constant in time. For $\mu<1$, all quantities display oscillatory behavior due to non-Markovian memory effects, with the amplitude of the oscillations gradually decreasing over time, as depicted in Figures \ref{Fig 8}(a-d).

	\onecolumngrid
	\begin{center}
		\begin{minipage}{0.48\textwidth}
			\centering
			\includegraphics[width=\linewidth]{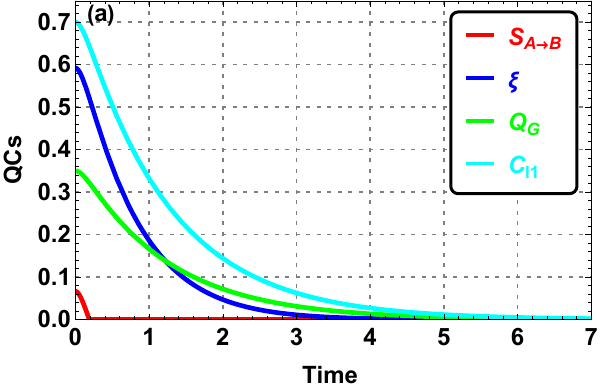}
		\end{minipage}
		\begin{minipage}{0.48\textwidth}
			\centering
			\includegraphics[width=\linewidth]{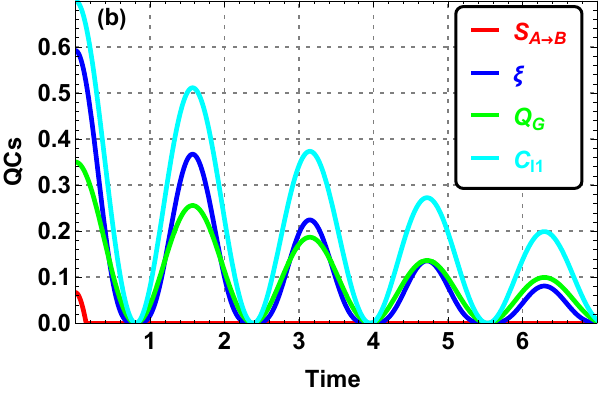}
		\end{minipage}
		\captionof{figure}{Comparison of quantum steering, entanglement of formation, geometric quantum discord, and quantum coherence in (a) the Markovian regime with $\tau = 0.1$ and (b) the non-Markovian regime with $\tau = 5$. The fixed parameters are $\Delta_0=-2$, $\omega = 3$, $\mu = 0$, and $T = 0.2$.	
		}\label{Fig 9}
	\end{center}
	\twocolumngrid
	As shown in Figure \ref{Fig 9}(a), the temporal evolution of quantum steering, entanglement, geometric quantum discord, and quantum coherence is examined in the Markovian regime.  The hierarchy among these quantifiers remains unchanged throughout the evolution, following the ordering: $\text{coherence} \supseteq \text{geometric quantum discord} \supseteq \text{entanglement} \supseteq \text{steerability}$.
	
	This demonstrates that the UdW may remain entangled even when steerability vanishes, and more generally, that entanglement does not necessarily imply steerability. In contrast, the geometric quantum discord ($Q_G$) and the coherence ($C_{11}$) prove to be the most robust, maintaining nonzero values well beyond the disappearance of steering and entanglement. This contrast highlights that the system retains its fundamental quantum character—namely, superposition and general nonclassical correlations—even after losing the resources that are most valuable for quantum communication and information-processing protocols, such as nonlocality and steerability.
	
	Figure \ref{Fig 9}(b) displays the evolution of the same correlations in the non-Markovian regime. A similar general behavior is observed; however, in this case, the correlations evolve sinusoidally and in phase, with oscillations whose amplitude gradually decreases. This indicates that non-Markovian memory effects play a significant role in shaping the dynamics of quantum correlations, leading to more coherent and periodic behavior compared to the Markovian regime, which is dominated by stronger decoherence processes.
	\section{Quantum thermodynamics} \label{sec:thermodynamics}
	The eigenvalues of $\rho_{AB}$ are straightforwardly obtained as 
	\begin{equation}
		\lambda_1=\rho_{11},\quad \lambda_2=\rho_{22},\quad  \lambda_3=\rho_{22}+\rho_{23},\quad \lambda_4=\rho_{22}-\rho_{23},
	\end{equation}
	which satisfy $\sum_{k=1}^{4} \lambda_k=1$.
	
	Hence, the von Neumann entropy is given by \cite{Albayrak2013} 
	\begin{equation}
		S=-\sum_{k=1}^{4}\lambda_k\ln\lambda_k
		\label{38}.
	\end{equation}
	The internal energy is then defined by 
	\begin{equation}
		U=\Tr\left[H_{\text{eff}}\rho_{AB} \right],
	\end{equation}
	in the weak-coupling limit, the internal energy simplifies to 
	\begin{equation}
		U=\omega\left(\rho_{11}-\rho_{22} \right).
	\end{equation}
	The quantum Stirling cycle (QSC) consists of four successive thermodynamic stages---two quantum isothermal processes and two isochoric processes, as illustrated in Figure \ref{CS}---where, during the isothermal transformations, the system remains in thermal equilibrium with a heat reservoir while the characteristic energy parameter ($\omega$) is varied, and in the isochoric stages, the system exchanges heat with the baths while $\omega$ remains constant; the four stages are described as follows \cite{Souf2025,BahaminPili2024}
	\begin{center}
		\includegraphics[width=0.87\linewidth]{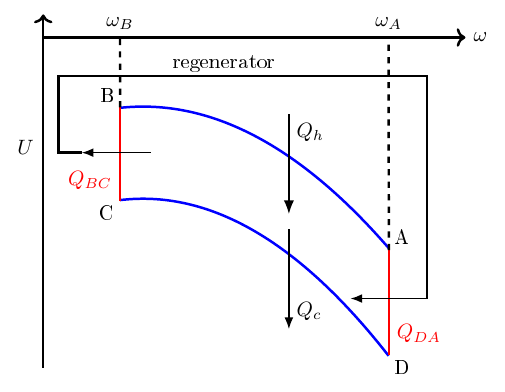}
		\captionof{figure}{The diagram of the Stirling cycle \cite{Huang2014}, where $Q_{h}$ denotes the absorbed heat, $Q_{c}$ the released heat, $T_{h}$ the hot-bath temperature, and $T_{c}$ the cold-bath temperature.}
		\label{CS}
	\end{center}
	
	\textbf{A$\to $B (Isothermal expansion):}
	The working substance interacts with a hot thermal bath at temperature $T_h$, while the energy parameter varies from. During this process, the system exchanges heat with the reservoir, which can be expressed as
	\begin{align}
		Q_{AB} = T_h \int_A^B dS = T_h(S_B - S_A),
	\end{align}
	where the entropy $S$ is determined from Eq.~(\ref{38}).
	
	\textbf{B$\rightarrow$C (Isochoric cooling):}
	The energy parameter $\omega$ remains constant at $\omega = \omega_B$, while the temperature decreases from $T_h$ to $T_c$. The exchanged heat is given by
	\begin{align}
		Q_{BC} = U_C - U_B = U(\omega_B, T_c) - U(\omega_B, T_h),
	\end{align}
	where $U$ denotes the internal energy of the system. 
	
	\textbf{C$\rightarrow$D (Isothermal compression):}
	The energy parameter $\omega$ is reduced from $\omega_B$ to $\omega_A$, while the temperature is maintained at $T_c$. The exchanged heat during this step is
	\begin{align}
		Q_{CD} = T_c \int_C^D dS = T_c(S_D - S_C).
	\end{align}
	
	\textbf{D$\rightarrow$A (Isochoric heating):}
	The energy parameter remains constant at $\omega=\omega_A$, while the temperature increases from $T_c$ to $T_h$. The absorbed heat is expressed as
	\begin{align}
		Q_{DA} = U_A - U_D = U(\omega_A, T_h) - U(\omega_A, T_c).
	\end{align}
	
	Over a complete cycle, the total absorbed and released  heats are given by
	\begin{align}
		Q_h = Q_{AB} + Q_{DA}, \qquad Q_c = Q_{BC} + Q_{CD},
	\end{align}
	where $Q_{AB}$ and $Q_{CD}$ correspond to the input and output heats, respectively. The total work performed by the system is
	\begin{equation}
		W = Q_{AB} + Q_{BC} + Q_{CD} + Q_{DA} ,
	\end{equation}
	and the efficiency $\eta$ of the quantum heat engine is expressed as
	\begin{equation}
		\eta = \frac{W}{Q_h}.
	\end{equation}
	\subsection{\bf Steady state}
	Figure \ref{FIG 11}(a) presents the behavior of the absorbed heat ($Q_{\text{h}}$), the released heat ($Q_{\text{c}}$), and the work output ($W$) as functions of $\omega_{B}$, for $\omega_{A}=1$, $T_{h}=2T_{c}$, and $\Delta_0=-1.5$. This figure shows that when $\omega_{B}$ is varied from $0$ to $1$, both $Q_{\text{h}}$ and $W$ maintain positive values. Conversely, $Q_{\text{c}}$ remains negative within the interval $0 < \omega_{B} < 0.9$. This sign convention confirms that the system operates as a heat engine in this specific regime.
	
	Figure \ref{FIG 11}(b) shows the efficiency ($\eta$) of the engine as a function of the energy spacing ($\omega_B$) for different initial spacings ($\omega_A$). The efficiency decreases as $\omega_B$ increases. This indicates that a larger energy spacing during the cold stroke reduces the work output relative to the heat absorbed during the hot stroke, consequently lowering the conversion efficiency. The rate of decrease depends on $\omega_A$: higher initial spacings correspond to a smaller efficiency loss for the same increase in $\omega_B$. In all cases, the efficiency remains below the Carnot limit $\eta_c = 1 - T_c/T_h$, 	as expected for an irreversible cycle. In summary, the figure illustrates that increasing the energy spacing in the cold branch negatively affects the engine performance, highlighting the sensitivity of quantum engines to the tuning of energy-level spacings.
	\onecolumngrid
	\begin{center}
		\vspace*{0.5cm}
		\begin{minipage}{0.48\textwidth}
			
			\centering
			\includegraphics[width=\linewidth]{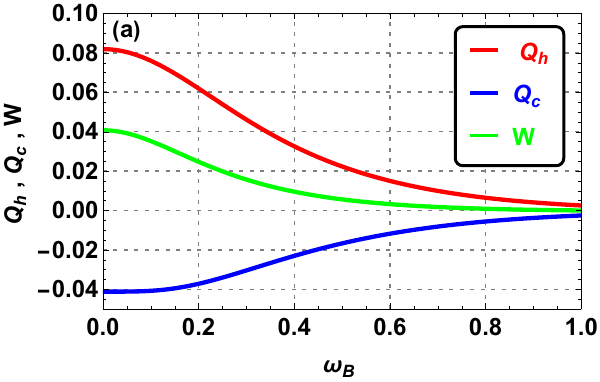}
		\end{minipage}
		\begin{minipage}{0.48\textwidth}
			\centering
			\includegraphics[width=\linewidth]{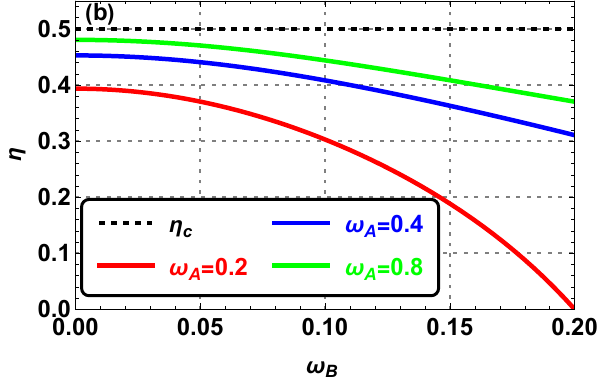}
		\end{minipage}
		\captionof{figure}{Variation of $Q_h$, $Q_c$, the work done $W$ (a) and the efficiency $\eta$ (b) versus $\omega_B$ with $\Delta_0=-1.5$, $T_h = 2T_c$, and $\omega_A=1$   
		}
		\label{FIG 11}
		\vspace*{0.3cm}
		\begin{minipage}{0.48\textwidth}
			\centering
			\includegraphics[width=\linewidth]{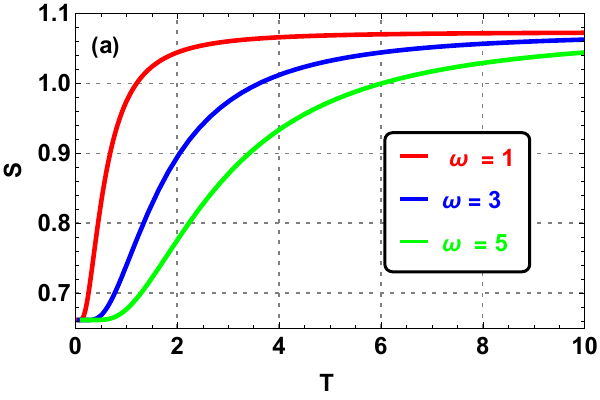}
		\end{minipage}
		\begin{minipage}{0.48\textwidth}
			\centering
			\includegraphics[width=\linewidth]{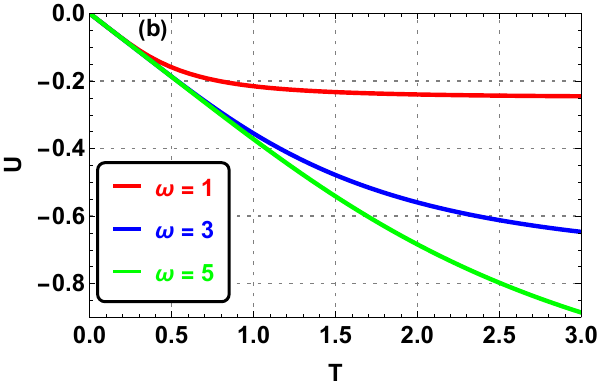}
		\end{minipage}
		\captionof{figure}{Variation of the (a) entropy, (b) internal energy versus $T$ for different values of the energy $\omega$ with $\Delta_0=-1.5$\label{FIG 12}
		}
	\end{center}
	\twocolumngrid
The von Neumann entropy S as a function of the bath temperature T for three values of the energy spacing $\omega$ is shown in
Fig.\ref{FIG 12}(a). For all cases, the entropy increases monotonically
with temperature, reflecting the progressive thermal mixing of
the system. Overall, the figure highlights the strong dependence
of entropy on both temperature and the energy-level spacing.
	
	Figure \ref{FIG 12}(b) illustrates the behavior of the internal energy ($U$) as a function of the bath temperature ($T$) for different values of spacing ($\omega$).  For all values of $\omega$, the internal energy remains negative and decreases monotonically with temperature. Overall, the figure shows that both $T$ and $\omega$ strongly influence the energetic response of the system, with lower energy spacing producing deeper negative internal-energy values.
	
	\onecolumngrid
	\vspace*{0.1cm}
	\begin{center}
		\begin{minipage}{0.48\textwidth}
			\centering
			\includegraphics[width=\linewidth]{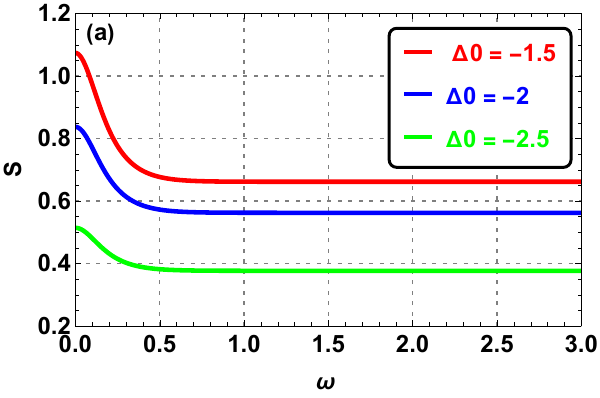}
		\end{minipage}
		\begin{minipage}{0.48\textwidth}
			\centering
			\includegraphics[width=\linewidth]{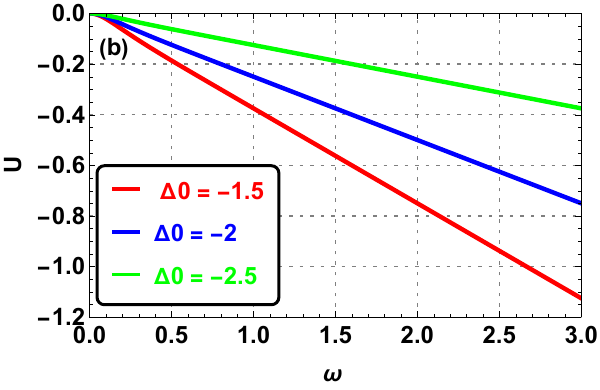}
		\end{minipage}
		\vspace*{-0.35cm}
		\captionof{figure}{Variation of the (a) entropy, (b) internal energy versus $\omega$ for different values of  the initial state parameter \(\Delta_0\) with $T=0.1$.}
		\label{FIG 13}
	\end{center}
	\twocolumngrid
	
	The behavior of the von Neumann entropy ($S$) as a function of the energy spacing ($\omega$) for various values of the initial state parameter ($\Delta_0$) is shown in Figure \ref{FIG 13}(a).  The entropy decreases rapidly for $\omega \lesssim 0.4$, indicating a fast reduction in the quantum mixedness of the system, and then remains nearly constant for $\omega \gtrsim 0.5$. Furthermore, larger absolute values of $\Delta_0$ correspond to generally lower entropy, demonstrating that the initial state parameter strongly affects the degree of quantum uncertainty. These results highlight the combined influence of the initial state configuration and energy spacing on the system's quantum correlations and thermal properties.
	
	Figure~\ref{FIG 13}(b) illustrates the behavior of the internal energy $U$ as a function of the energy spacing $\omega$ for different values of the initial state parameter $\Delta_0 $ at a fixed temperature $T = 0.1$. For all considered values of $\Delta_0$, the internal energy is negative and decreases linearly with $\omega$. Furthermore, larger absolute values  of $\Delta_0$ correspond to generally higher internal energy, demonstrating that the initial state strongly affects the system’s energy content. These results highlight the combined effect of the initial state configuration and energy spacing on the thermodynamic properties of the system at this temperature.
	\subsection{\bf Dynamical state}
	\onecolumngrid
	\begin{center}
		\vspace*{0.4cm}
		\begin{minipage}{0.48\textwidth}
			\centering
			\includegraphics[width=\linewidth]{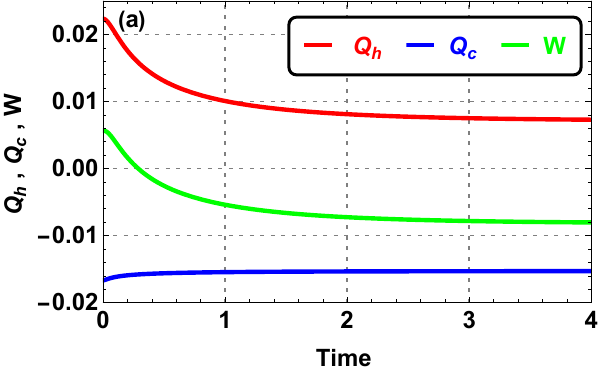}
		\end{minipage}
		\begin{minipage}{0.48\textwidth}
			\centering
			\includegraphics[width=\linewidth]{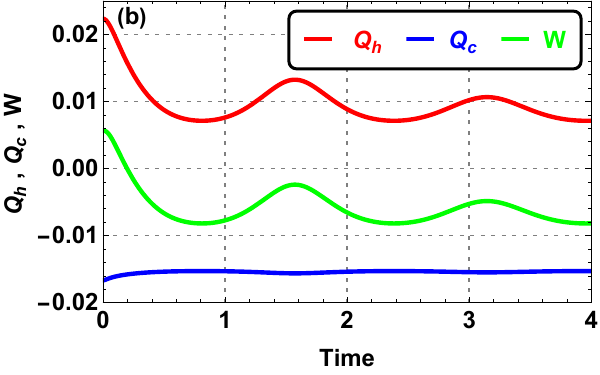}
		\end{minipage}
		\captionof{figure}{Comparison of $Q_h$, $Q_c$, the work done $W$ in (a) the Markovian regime with $\tau = 0.1$ and (b) the non-Markovian regime with $\tau = 5$. The fixed parameters are $\Delta_0=-1.5$, $\omega_A = 1$, $\omega_B = 0.5$, $\mu = 0.4$, and $T_h=2T_c$.	
			\label{FIG 14}}		
	\end{center}
	\twocolumngrid 
	Figure \ref{FIG 14} compares the heat exchanged with the hot bath ($Q_h$), the heat released to the cold bath ($Q_c$), and the extracted work ($W$) as functions of time in the Markovian ($\tau=0.1$) and non-Markovian ($\tau=5$) regimes. In the Markovian case, all quantities evolve monotonically toward their steady values, indicating a smooth, irreversible relaxation. In contrast, the non-Markovian dynamics exhibits pronounced oscillations in $Q_h$, $Q_c$, and $W$, reflecting memory-induced backflow of information and energy between the working substance and the reservoir. These oscillations enhance the instantaneous work output but delay the stabilization of the thermodynamic cycle. Overall, the figure demonstrates that non-Markovian memory effects significantly modify the energetic exchanges and transient work production of the quantum heat engine.
	
	\onecolumngrid
	\begin{center}
		\begin{minipage}{0.48\textwidth}
			\centering
			\includegraphics[width=\linewidth]{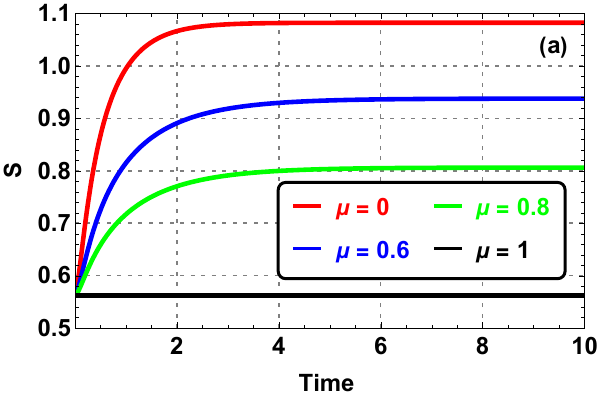}
		\end{minipage}
		\begin{minipage}{0.48\textwidth}
			\centering
			\includegraphics[width=\linewidth]{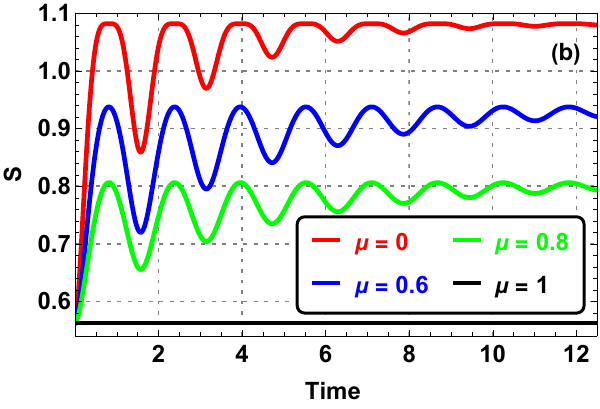}
		\end{minipage}
		\captionof{figure}{Comparison of the entropy ($S$) in (a) the Markovian regime ($\tau = 0.1$) and (b) the non-Markovian regime ($\tau = 5$) for different values of the classical correlation parameter ($\mu$). The fixed parameters are $\Delta_0=-2$, $\omega = 2$, and $T=0.1$.}
		\label{FIG 15}
	\end{center}
	\twocolumngrid
	Figure~\ref{FIG 15} compares the time evolution of the von Neumann 
	entropy $S$ for different values of the classical correlation parameter 
	$\mu$ in both (a) the Markovian regime ($\tau = 0.1$) and 
	(b) the non-Markovian regime ($\tau = 5$). In the Markovian case, the entropy increases monotonically to achieve its steady-state, it is observed that the entropy grows over time, whereas increasing values of $\mu$ lead to a reduction in entropy.
	
	In contrast, the non-Markovian regime exhibits pronounced oscillations reflecting memory effects. Although the amplitude of these revivals becomes more significant as $\mu$ decreases, it gradually decreases over time.
	\onecolumngrid
	\begin{center}
		\begin{minipage}{0.48\textwidth}
			\centering
			\includegraphics[width=\linewidth]{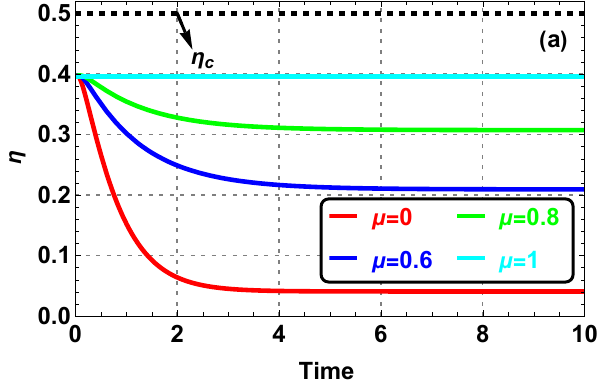}
		\end{minipage}
		\begin{minipage}{0.48\textwidth}
			\centering
			\includegraphics[width=\linewidth]{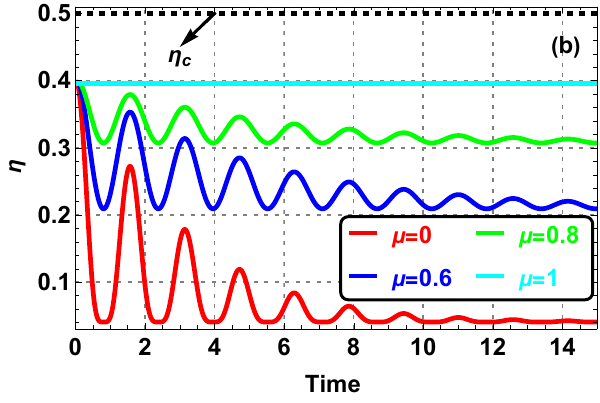}
		\end{minipage}
		\captionof{figure}{Comparison of the efficiency ($\eta$) in (a) the Markovian regime ($\tau = 0.1$) and (b) the non-Markovian regime ($\tau = 5$) for different values of the classical correlation parameter ($\mu$).  The fixed parameters are $\Delta_0=-1.5$, $\omega_A = 1$, $\omega_B = 0.5$, and $T_h=2T_c$.}
		\label{FIG 16}
	\end{center}
	\twocolumngrid
	
	Figure \ref{FIG 16} compares the efficiency ($\eta$) of the quantum heat engine in the (a) Markovian regime ($\tau = 0.1$) and (b) non-Markovian regime ($\tau = 5$) across several values of the classical correlation parameter ($\mu$). In both regimes, the efficiency increases with stronger classical correlations ($\mu$), indicating that such correlations enhance the engine's performance. However, the efficiency gradually decreases over time in both regimes. In the non-Markovian regime ($\tau=5$), more pronounced oscillations and a higher peak efficiency are observed. These oscillations reflect the backflow of information that helps preserve quantum coherence during the cycle, although their amplitude gradually decreases over time. Overall, memory effects significantly improve the engine's performance compared to the purely Markovian regime ($\tau=0.1$).
	
	\onecolumngrid
	\begin{center}
		\vspace*{0.1cm}
		\begin{minipage}{0.48\textwidth}
			\centering
			\includegraphics[width=\linewidth]{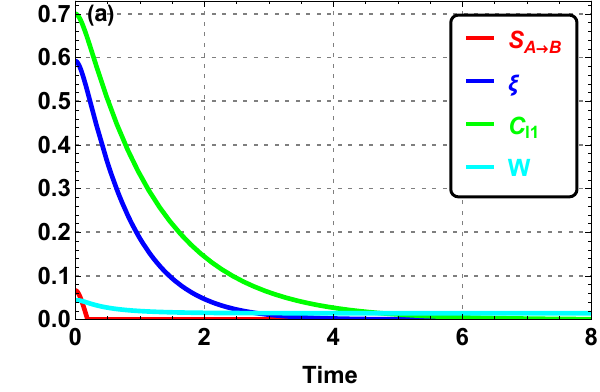}
		\end{minipage}
		\begin{minipage}{0.48\textwidth}
			\centering
			\includegraphics[width=\linewidth]{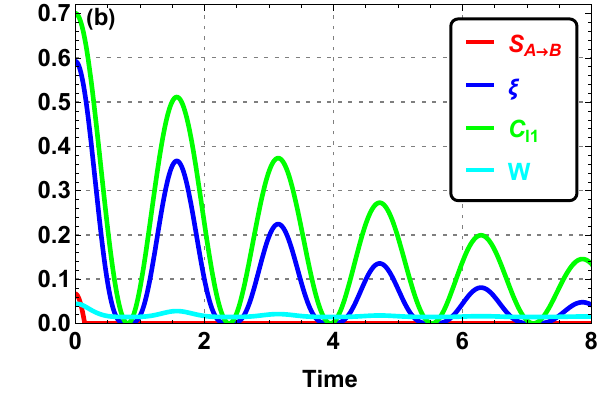}
		\end{minipage}
		\captionof{figure}{Comparison of quantum steering, entanglement of formation, quantum coherence, and the work done  in (a) the Markovian regime with $\tau = 0.1$ and in (b) the non-Markovian regime with $\tau = 5$. The fixed parameters are $\Delta_0=-1.8$, $\omega = 3$, $\mu = 0$, $T = 0.2$, $\omega_A=2$, $\omega_B=0.5$ and $T_h=1.5T_c$ .	
		}
		\label{FIG 17}
	\end{center}
	\twocolumngrid
	
	As illustrated in Figure \ref{FIG 17}, the time evolution of quantum steering ($S_{A\to B}$), entanglement ($\xi$), quantum coherence ($C_{l1}$), and the extracted work ($W$) is analyzed for both the Markovian ($\tau=0.1$) and non-Markovian ($\tau=5$) regimes. In the Markovian regime, steering rapidly vanishes, the work remains relatively weak, entanglement gradually decreases, and quantum coherence persists longer before fading. In the non-Markovian regime, more pronounced oscillations are observed for all quantities, reflecting memory effects and information backflow, yet their amplitude also gradually decreases over time. We observe that an increase in coherence results in a higher amount of extracted work. Conversely, the extracted work decreases as the bath temperature rises. Moreover, for $t \gtrsim 4$, these quantities gradually vanish over time due to the influence of the environment.
	Figure \ref{FIG 17} shows that the extracted work, quantum steering and entanglement are bounded by coherence. Moreover,
	entanglement and coherence, as quantum resources, are beneficial for work extraction.
	\section{FEASIBILITY}	\label{sec:FEASIBILITY}

 The direct experimental verification of the Unruh effect remains extremely challenging due to the enormous accelerations required to produce a measurable temperature. Indeed, an Unruh temperature of only a few Kelvin would require accelerations on the order of $10^{20}\,\mathrm{m/s^2}$ \cite{Crispino2008}. Such accelerations are currently beyond the reach of conventional laboratory setups. Nevertheless, several experimental and analog proposals have been put forward to probe acceleration-induced thermal phenomena and related effects.
	
	One prominent direction involves ultra-intense laser–electron interaction schemes, where electrons are subjected to extremely strong electromagnetic fields, potentially reaching accelerations at which Unruh-like signatures may become observable \cite{Chen1999, Schutzhold2006}. Another line of investigation considers electrons in storage rings, where circular acceleration may lead to observable effects related to acceleration-induced vacuum fluctuations \cite{Bell1983}. Although such scenarios do not provide a direct measurement of the Unruh temperature, they offer experimentally accessible regimes in which acceleration-induced modifications of quantum states can be studied.
	
In parallel, analog gravity platforms have emerged as promising testbeds for simulating Unruh-type phenomena. Systems such as Bose–Einstein condensates, superconducting circuits, and optical cavity setups allow effective horizons or accelerated detector models to be engineered in controlled laboratory environments \cite{Garay2000, Fuentes2005}. These platforms enable the investigation of detector–field interactions and thermal-like responses without requiring physically unattainable accelerations.
	
	From the open quantum systems perspective adopted in this work, the Unruh effect manifests as an effective thermal environment perceived by the accelerating detectors. In this framework, the scalar field vacuum plays the role of a structured quantum bath satisfying the Kubo--Martin--Schwinger (KMS) condition. The resulting dynamics, described by a Kossakowski--Lindblad master equation, incorporate both decoherence and dissipation processes induced by the field. This viewpoint highlights that the Unruh effect is not merely a particle-detection phenomenon but also a mechanism for environment-induced quantum correlations.
	
	Importantly, recent theoretical proposals suggest that, instead of attempting to measure the Unruh temperature directly, one may look for indirect signatures in quantum resources—such as entanglement or coherence—generated through field-mediated correlations between detectors \cite{Reznik2003, Benatti2004}. Such resource-based signatures may provide more experimentally accessible indicators of acceleration-induced vacuum fluctuations, particularly in analog or engineered quantum platforms.
	
	Therefore, while a direct observation of the Unruh effect in its original form remains elusive, ongoing advances in high-intensity laser physics, quantum simulation, and analog gravity systems suggest that indirect experimental probes of acceleration-induced quantum phenomena may become feasible in the foreseeable future.
	
	\section{Conclusion}
	\label{sec:conclusion}
In this work, we have explored the intricate relationship between quantum correlations, quantum coherence and quantum thermodynamics in a quantum heat engine modeled by Unruh--DeWitt detectors. By examining both the steady and dynamical states of quantum resources---such as entanglement, geometric quantum discord, and coherence---we showed that their evolution is profoundly influenced by the characteristics of the surrounding environment. In particular, we analyzed the behavior of coherence and various forms of quantum correlations with respect to the temperature ($T$), the energy spacing ($\omega$), and the initial state parameter ($\Delta_0$). This allowed us to compare not only coherence with different measures of quantum correlations, but also the interplay between these quantum resources and the thermodynamic performance of the engine. Our analysis reveals that non-Markovian memory effects play a constructive role: they mitigate the degradation of quantum resources and enhance key thermodynamic quantities, including heat exchange, work extraction, and overall engine efficiency. These findings underscore the central importance of both quantum and classical correlations in determining the operational performance of quantum thermal devices. Beyond providing deeper insight into the connections between quantum information and thermodynamics, our study highlights practical avenues for optimizing quantum technologies operating in relativistic or open-system regimes. Future research may extend this framework to multipartite systems, more general detector trajectories, or strong-coupling regimes, potentially uncovering additional mechanisms through which environmental memory and quantum resources can be harnessed to boost the performance of quantum machines.%
	
	\section*{Acknowledgments}

The authors extend their appreciation to the Deanship of Research and Graduate Studies at King Khalid University for funding this work through Large Research Project under grant number RGP2/172/46.
		
	\appendix %
	\label{App:A}

	\section{Derivation of the Kossakowski Coefficients}
The derivation of the Kossakowski matrix follows the standard
weak-coupling Born–Markov procedure leading to the
Gorini–Kossakowski–Lindblad–Sudarshan (GKLS) master equation
\cite{Gorini1976,Lindblad1976,BenattiFloreanini2004}.
	\subsection{ Interaction Picture and Born Approximation}
	We consider the total Hamiltonian
	\begin{equation}
		H = H_0 + H_{\phi} + H_I ,
	\end{equation}
	where the interaction term is
	\begin{equation}
		H_I = \mu \left[
		(\sigma_2^{(A)} \otimes I^{(B)}) \phi(x_1)
		+
		(I^{(A)} \otimes \sigma_2^{(B)}) \phi(x_2)
		\right].
	\end{equation}
	In the interaction picture, the total density matrix evolves as
	\begin{equation}
		\frac{d\rho_{\text{tot}}(t)}{dt}
		=
		-i [H_I(t), \rho_{\text{tot}}(t)] .
	\end{equation}
	Assuming weak coupling ($\mu \ll 1$) and an initially separable
	state
	\begin{equation}
		\rho_{\text{tot}}(0) = \rho_{AB}(0) \otimes |0\rangle\langle 0|,
	\end{equation}
	we apply the Born approximation,
	\begin{equation}
		\rho_{\text{tot}}(t)
		\approx
		\rho_{AB}(t) \otimes |0\rangle\langle 0|.
	\end{equation}
	\subsection{ Markov Approximation and Reduced Dynamics}
	Tracing over the field degrees of freedom and performing the
	Markov approximation, we obtain the reduced master equation
	\begin{equation}
		\frac{d\rho_{AB}}{dt}
		=
		- \int_0^{\infty} d\tau
		\,
		\text{Tr}_{\phi}
		\left[
		H_I(t),
		\left[
		H_I(t-\tau),
		\rho_{AB}(t) \otimes |0\rangle\langle 0|
		\right]
		\right].
	\end{equation}
	The field correlations enter through the Wightman function
	\begin{equation}
		Y^{+}(x,x')
		=
		\langle 0| \phi(x) \phi(x') |0\rangle .
	\end{equation}
	Along the detector trajectory, this reduces to $Y^{+}(\tau)$.
	\subsection{ Fourier Transform and Dissipative Coefficients}
	The dissipative part of the master equation depends on the
	Fourier transform of the Wightman function defined in Eq.~(8),
	\begin{equation}
		Y(\lambda)
		=
		\int_{-\infty}^{+\infty}
		d\tau \,
		e^{i\lambda\tau}
		Y^{+}(\tau).
	\end{equation}
	Evaluating the integrals yields the coefficients
	\begin{equation}
		\gamma_{\pm}
		=
		Y(\omega) \pm Y(-\omega),
	\end{equation}
	and
	\begin{equation}
		\gamma_0
		=
		Y(0) - \frac{\gamma_+}{2}.
	\end{equation}
	\subsection{ Construction of the Kossakowski Matrix}
	Collecting all contributions and performing the standard
	Gorini–Kossakowski–Lindblad–Sudarshan decomposition,
	the master equation takes the form
	\begin{equation}
		\frac{d\rho_{AB}}{dt}
		=
		-i[H_{\text{eff}}, \rho_{AB}]
		+
		\sum_{i,j=1}^{3}
		C_{ij}
		\left(
		\sigma_j \rho_{AB} \sigma_i
		-
		\frac{1}{2}
		\{\sigma_i \sigma_j, \rho_{AB}\}
		\right).
	\end{equation}
		Using the algebra of Pauli matrices,
	\begin{equation}
		\sigma_i\sigma_j
		=
		\delta_{ij}I + i\epsilon_{ijk}\sigma_k,
	\end{equation}
	one obtains the Kossakowski matrix
	\begin{equation}
		C_{ij}
		=
		\frac{\gamma_+}{2}\delta_{ij}
		-
		i\frac{\gamma_-}{2}\epsilon_{ijk}\delta_{3k}
		+
		\gamma_0 \delta_{3i}\delta_{3j},
	\end{equation}
	This shows explicitly how the coefficients $C_{ij}$
	originate from the field correlation functions evaluated
	along the accelerated trajectories. The thermal properties
	induced by acceleration are encoded in the Fourier
	components $Y(\pm\omega)$, which determine $\gamma_\pm$
	and therefore the dissipative structure of the reduced dynamics.

	\bibliographystyle{apsrev4-1}
	\bibliography{references}
\end{document}